\documentclass[epj]{svjour}
\usepackage{graphics}
\usepackage{amsmath}
\usepackage{amssymb}
\usepackage{amsfonts}
\usepackage{graphicx}
\usepackage{color}
\begin{document}
%%%%%%%%%%%%%%%%%%%%%%%%%%%%%%%%%%%%%%%%%%%%%%%%%%%%%%%%%%%%%%%%%%%%%%%%%%%%%%%%%%%%%%%%%%
\title{Chimera states and synchronization in magnetically driven SQUID metamaterials}
%%%%%%%%%%%%%%%%%%%%%%%%%%%%%%%%%%%%%%%%%%%%%%%%%%%%%%%%%%%%%%%%%%%%%%%%%%%%%%%%%%%%%%%%%%
\abstract{
One-dimensional arrays of Superconducting QUantum Interference Devices (SQUIDs) form
magnetic metamaterials exhibiting extraordinary properties, including tunability, 
dynamic multistability, negative magnetic permeability, and broadband transparency. 
The SQUIDs in a metamaterial
interact through non-local, magnetic dipole-dipole forces, that makes it possible for
multiheaded chimera states and coexisting patterns, including solitary states, to appear.
The spontaneous emergence of chimera states and the role of multistability
is demonstrated numerically for a SQUID metamaterial driven by an alternating magnetic 
field. The spatial synchronization and temporal complexity are discussed and the
parameter space for the global synchronization reveals the areas of coherence-incoherence
transition. Given that both one- and two-dimensional SQUID metamaterials
have been already fabricated and investigated in the lab, the presence of a chimera state 
could in principle be detected with presently available experimental set-ups.} 
%%%%%%%%%%%%%%%%%%%%%%%%%%%%%%%%%%%%%%%%%%%%%%%%%%%%%%%%%%%%%%%%%%%%%%%%%%%%%%%%%%%%%%%%%%

%%%%%%%%%%%%%%%%%%%%%%%%%%%%%%%%%%%%%%%%%%%%%%%%%%%%%%%%%%%%%%%%%%%%%%%%%%%%%%%%%%%%%%%%%%
\author{
J. Hizanidis\inst{1} \and N. Lazarides\inst{1,2,3} \and G. Neofotistos\inst{1,3} \and 
G. P. Tsironis\inst{1,2,3}
}
\institute{
\inst{1}Crete Center for Quantum Complexity and Nanotechnology, Department of Physics,
      University of Crete, P. O. Box 2208, 71003 Heraklion, Greece. \\
\inst{2}Institute of Electronic Structure and Laser,
      Foundation for Research and Technology--Hellas, P.O. Box 1527, 71110 Heraklion,
      Greece. \\
\inst{3}National University of Science and Technology MISiS, Leninsky prosp. 4, Moscow,
      119049, Russia
}
%%%%%%%%%%%%%%%%%%%%%%%%%%%%%%%%%%%%%%%%%%%%%%%%%%%%%%%%%%%%%%%%%%%%%%%%%%%%%%%%%%%%%%%%%%
%%% \pacs{05.65.+b,05.45.Xt,78.67.Pt,89.75.-k,89.75.Kd}
%%%%%%%%%%%%%%%%%%%%%%%%%%%%%%%%%%%%%%%%%%%%%%%%%%%%%%%%%%%%%%%%%%%%%%%%%%%%%%%%%%%%%%%%%%
\maketitle
%%%%%%%%%%%%%%%%%%%%%%%%%%%%%%%%%%%%%%%%%%%%%%%%%%%%%%%%%%%%%%%%%%%%%%%%%%%%%%%%%%%%%%%%%%
\section{Introduction} 
\label{sec:sec1}
\begin{figure}[h!]
\includegraphics[angle=0, width=0.85 \linewidth]{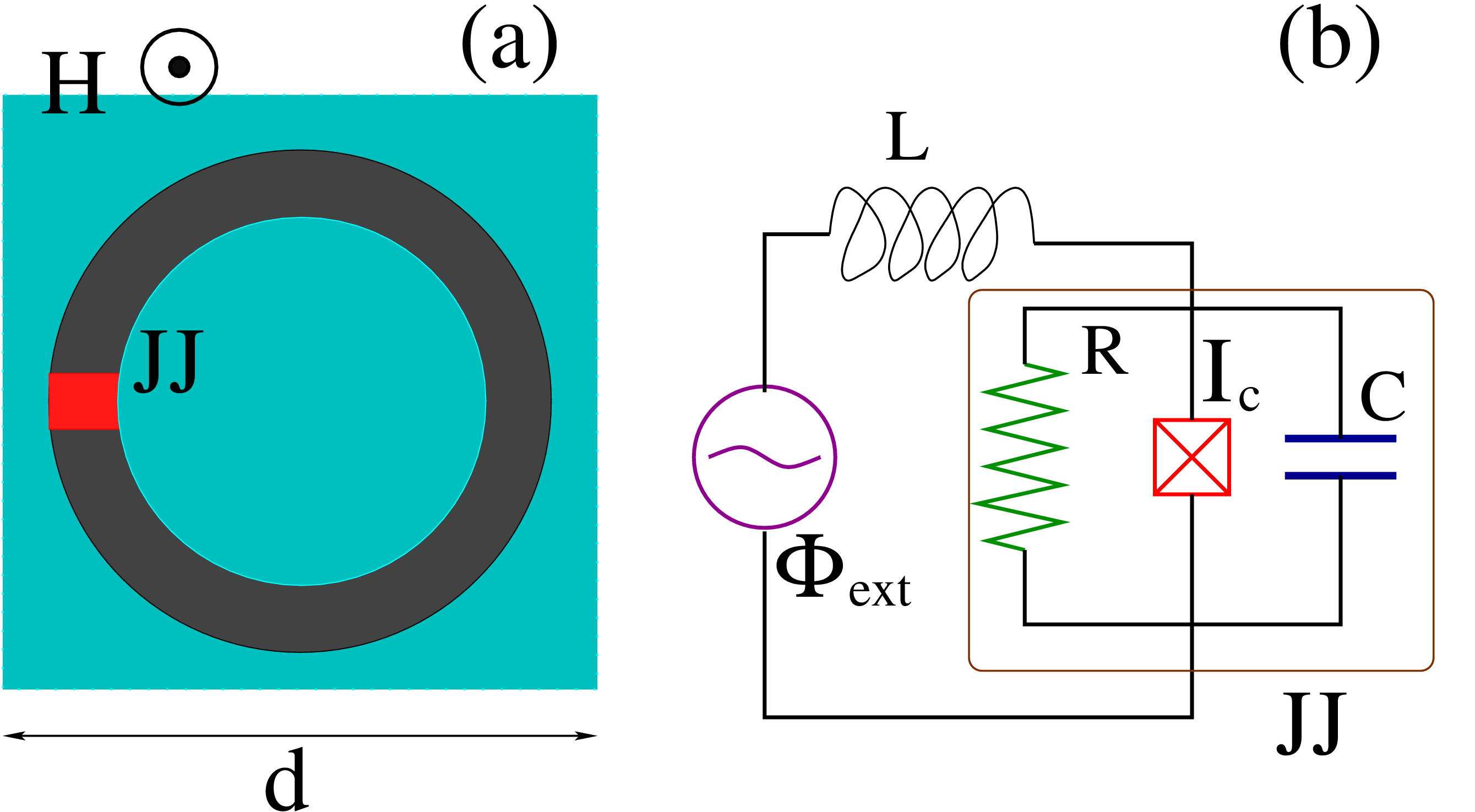} 
\includegraphics[angle=0, width=0.88 \linewidth]{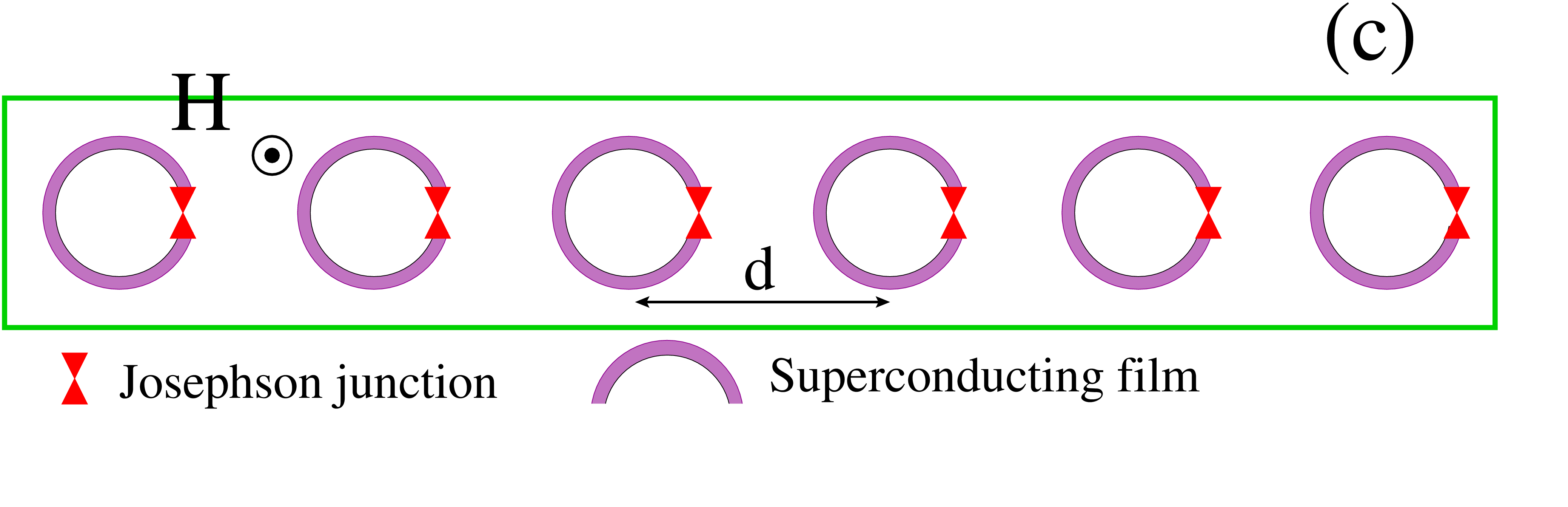} 
\caption{(Color online)
Schematic of an rf SQUID in an alternating magnetic field ${\bf H}(t)$ (a), and equivalent 
electrical circuit (b). The real Josephson junction is represented by the circuit elements
that are in the brown-dashed box. (c) A one-dimensional SQUID metamaterial is formed by 
repetition of the squared unit cell of side $d$ shown in (a). 
\label{fig1}
}
\end{figure}

Superconducting QUantum Interference Device (SQUID) metamaterials are superconducting
artificial media whose function relies both on their geometry and the extraordinary
properties of superconductivity and the Josephson effect \cite{Anlage2011,Jung2014}.
Recent experiments on one- and two-dimensional radio-frequency (rf) SQUID metamaterials 
have revealed their wide-band tuneability, significantly reduced losses, dynamic
multistability, and tunable broadband transpare- ncy 
\cite{Jung2014,Jung2013,Butz2013a,Trepanier2013,Jung2014b,Zhang2015}. The simplest 
version of an rf SQUID involves a superconducting ring interrupted by a Josephson junction 
\cite{Josephson1962} (Fig. \ref{fig1}a); this device is a highly nonlinear resonator with 
a strong response to applied magnetic fields. SQUID metamaterials exhibit peculiar magnetic
properties such as negative diamagnetic permeability, predicted both for the quantum 
\cite{Du2006} and the classical \cite{Lazarides2007} regime. The applied alternating
fields induce (super)currents in the SQUID rings, which are therefore coupled through
dipole-dipole magnetic forces. This interaction is weak due to its magnetic nature.
However, it couples the SQUIDs non-locally since it falls-off as the inverse cube of their
center-to-center distance.

Nonlocal coupling in systems of interacting oscillators is capable of producing 
counter-intuitive space-time patterns, called chimera states\cite{KUR02a,ABR04}. Chimera
states are characterized by the coexistence of synchronous and asynchronous clusters of 
oscillators. They were first observed over a decade ago, in systems of identical phase
oscillators with  symmetric  coupling \cite{KUR02a,ABR04}, and have been intensively
studied both theoretically
\cite{Omelchenko2008,Abrams2008,Pikovsky2008b,Ott2009,Martens2010,Omelchenko2011,Yao2013,Omelchenko2013,vuellings:2014,Hizanidis2014,Zakharova2014,Bountis2014,Yeldesbay2014,hizanidis:2015}
and experimentally, in chemical \cite{tinsley:2012}, optical \cite{hagerstrom:2012},
mechanical \cite{martens:2013}, electronic \cite{larger:2013}, and electrochemical
\cite{schmidt:2014,wickramasinghe:2013} oscillator systems (for further reading refer 
to \cite{panaggio:2015}). Recent works also report on the issue of robustness of chimera
states \cite{omelchenko:2015} as well as on the emergence of chimera states in systems
with global \cite{BOE15} and local coupling schemes \cite{LAI15}. Chimera-like states
in modular networks \cite{SHA10,HIZ15} have also been investigated, expanding our
understanding on the role of topology and dynamics for their occurrence. Recent research
efforts aim to stabilize chimera states by feedback schemes \cite{SIE14} and to control 
the localization of the different regimes \cite{BIC15,ISE15}. In \cite{LAZ15}, chimera 
states with very long life-times were numerically demonstrated in SQUID metamaterials
with weak dissipation and non-local magnetic interactions due to an alternating external 
magnetic field. The SQUID metamaterial model had been previously used in the weak and
nearest neighbor coupling regime, for the  investigation of intrinsic localization and
tuneability effects \cite{Lazarides2008a,Lazarides2012,Lazarides2013b}. Moreover, the 
synchronization and metastability levels of the chimera states were discussed in terms 
of appropriate measures \cite{LAZ15}.
%%%%%%%%%%%%%%%%%%%%%%%%%%%%%%figure1%%%%%%%%%%%%%%%%%%%%%%%%%%%%%%%%%%%%%%%%%%%%%%%%%%%%%
%%%%%%%%%%%%%%%%%%%%%%%%%%%%%%endfigure1%%%%%%%%%%%%%%%%%%%%%%%%%%%%%%%%%%%%%%%%%%%%%%%%%%

In the present work we extend this study and focus on the multistability of the system 
where different patterns coexist. Chimera states with single and multiple coherent 
regions of different sizes as well as other patterns will be demonstrated. In addition, 
by quantifying the spatial synchronization in the SQUID array we reveal the detailed 
structure of coherence and incoherence in the device. Finally, the parameter space for 
global synchronization transitions will be analyzed, providing insight into the role of 
the external driving field. Our results will help to provide further understanding to the 
implications of the second derivative term (inertia) in the occurrence of chimera states
\cite{Bountis2014,LAZ15,Jaros2015,Olmi2015}, which has not been sufficiently studied. 

In the next Section, we present the non-locally coupled SQUID metamaterial model and the 
measures for quantifying spatial and global synchronization. Current amplitude-driving 
frequency diagrams for a single SQUID which reveal its multistability property in the 
strongly nonlinear regime is also presented. In Section~\ref{sec:sec3}, 
the numerically obtained space-time evolution of several coexisting patterns are shown for 
different initial conditions. The magnitude of the 
spatial synchronization parameter is monitored in time for the non-locally coupled SQUID 
metamaterials and the temporal complexity is discussed in terms of phase portraits of the 
individual SQUID oscillators. In Section~\ref{sec:sec4}, the possibility that the SQUID 
metamaterial may undergo a synchronization-desynchronization transition is explored. 
Appropriate density plots mapping the order parameter on the dc flux bias - driving 
period plane unveil the relevant parameter regions in which such a transition may occur.
In the concluding section (Section~\ref{sec:sec5}) we summarize our results.

%%%%%%%%%%%%%%%%%%%%%%%%%%%%%%%%%%%%%%%%%%%%%%%%%%%%%%%%%%%%%%%%%%%%%%%%%%%%%%%%%%%%%%%%%%
\section{SQUID Metamaterial Modelling and Measure of Synchronization}
\label{sec:sec2}
The model under consideration involves a one-dimensional linear array of $N$ identical 
rf SQUIDs coupled together magnetically through dipole-dipole forces. The magnetic flux 
$\Phi_n$ threading the $n-$th SQUID loop is given by the following equation:
\begin{eqnarray}
\label{01}
  \Phi_n =\Phi_{ext} +L\, I_n +L\, \sum_{m\neq n} \lambda_{|m-n|} I_m ,
\end{eqnarray}
where the indices $n$ and $m$ run from $1$ to $N$, $\Phi_{ext}$ is the external flux in 
each SQUID, $\lambda_{|m-n|} =M_{|m-n|}/L$ is the dimensionless coupling coefficient 
between the SQUIDs at positions $m$ and $n$, with $M_{|m-n|}$ being their corresponding 
mutual inductance, and
\begin{eqnarray}
\label{02}
    -I_n =C\frac{d^2\Phi_n}{dt^2} +\frac{1}{R} \frac{d\Phi_n}{dt} 
                                +I_c\, \sin\left(2\pi\frac{\Phi_n}{\Phi_0}\right) 
\end{eqnarray}
is the current in the $n-$th SQUID given by the resistively and capacitively shunted 
junction (RCSJ) model \cite{Likharev1986}, with  $\Phi_0$ and $I_c$ being the flux quantum 
and the critical current of the Josephson junctions, respectively.  Within the RCSJ 
framework, $R$, $C$, and $L$ are the resistance, capacitance, and self-inductance of the 
SQUIDs' equivalent circuit (Fig. \ref{fig1}b). From Eqs. (\ref{01}) and (\ref{02}) we obtain 
the equation:
\begin{eqnarray}
\label{05}
  C\frac{d^2\Phi_n}{dt^2} +\frac{1}{R} \frac{d\Phi_n}{dt}
    +\frac{1}{L} \sum_{m=1}^N  \left( {\bf \hat{\Lambda}}^{-1} \right)_{nm} 
         \left( \Phi_m -\Phi_{ext} \right) 
   \nonumber \\
   +I_c\, \sin\left(2\pi\frac{\Phi_n}{\Phi_0}\right) =0 ,
\end{eqnarray}
where ${\bf \hat{\Lambda}}^{-1}$ is the inverse of the $N\times N$ coupling matrix 
\begin{eqnarray}
\label{04}
  {\bf \hat{\Lambda}} = \left\{ \begin{array}{ll}
        1, & \mbox{if $m= n$};\\
        \lambda_{|m-n|} =\lambda_0 \, |m-n|^{-3}, & \mbox{if $m\neq n$},\end{array} \right.   
\end{eqnarray}
with $\lambda_0$ being the coupling coefficient between nearest neighboring SQUIDs.

%%%%%%%%%%%%%%%%%%%%%%%%%%%%%%%%%%%%%%%%%%%%%%%%%%%%%%%%%%%%%%%%%%%%%%%%%%%%%%%%%%%%%%%%%%
Normalizing the frequency and time to $\omega_0 =1/\sqrt{LC}$ and its inverse 
$\omega_0^{-1}$, respectively, and the fluxes and currents to $\Phi_0$ and $I_c$, 
respectively, Eq. (\ref{05}) reads ($n=1,...,N$)
\begin{eqnarray}
\label{06}
  \ddot{\phi}_n +\gamma \dot{\phi}_n +\beta \sin\left( 2\pi \phi_n \right) 
    =\sum_{m=1}^N \left( {\bf \hat{\Lambda}}^{-1} \right)_{nm} 
         \left( \phi_{ext} -\phi_m \right) ,
\end{eqnarray}
where the overdots denote derivation with respect to the normalized temporal variable, 
$\tau$, $\phi_{ext} =\phi_{dc}+\phi_{ac} \cos(\Omega \tau )$, with $\phi_{dc}$ being the
dc flux component, $\Omega=\omega/\omega_0$ the normalized driving frequency, and 
\begin{equation}
\label{14}
   \beta=\frac{I_c L}{\Phi_0} =\frac{\beta_L}{2\pi}, \qquad
   \gamma=\frac{1}{R} \sqrt{ \frac{L}{C} } , 
\end{equation}
is the SQUID parameter and loss coefficient, respectively.

\begin{figure}[h!]
\includegraphics[clip,width=0.48\textwidth,angle=0]{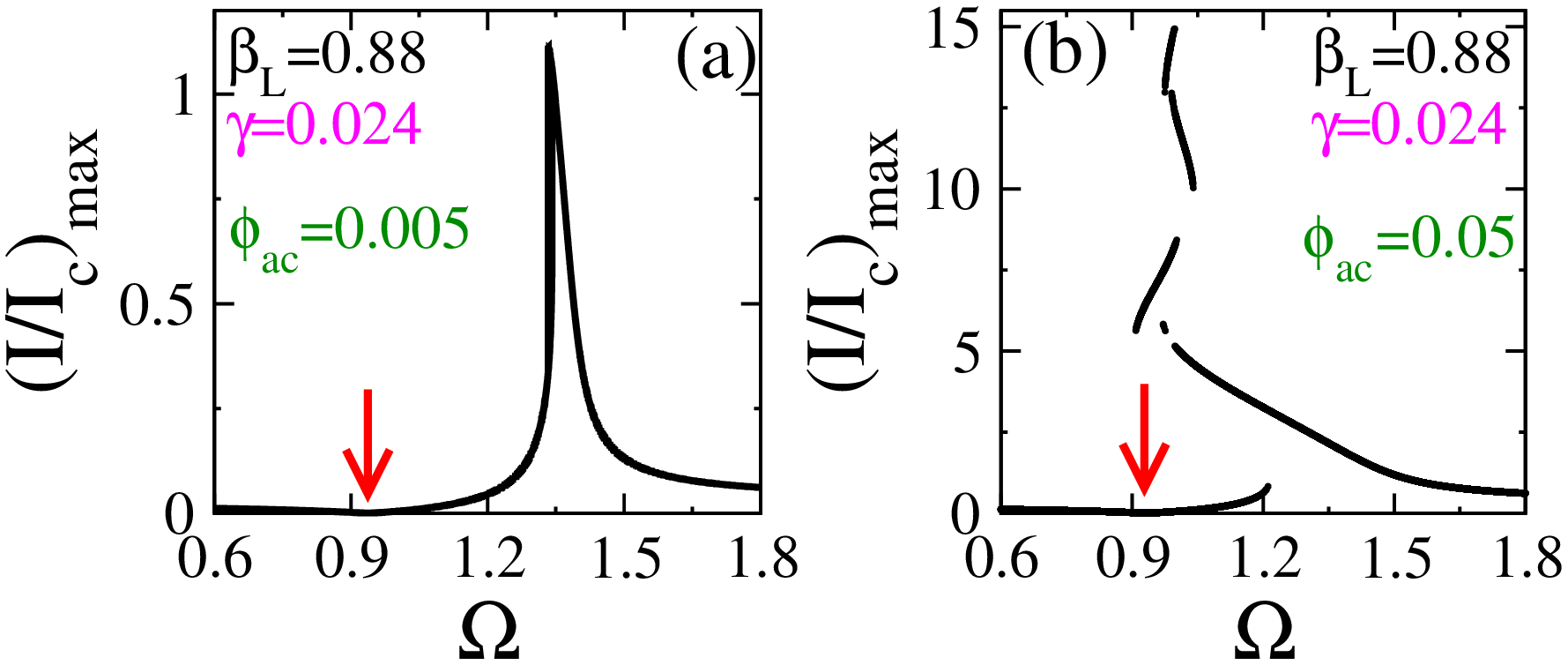}
\caption{(Color online)
Current amplitude $(I/I_c)_{max}$ - driving frequency $\Omega$ characteristics for a single
rf SQUID with $\beta_L =0.88$, $\gamma=0.024$, $\phi_{dc}=0$, and (a) $\phi_{ac}=0.005$;
(b) $\phi_{ac}=0.05$. The red arrows indicate the position of an anti-resonance (see text).
\label{fig999} }
\end{figure}

\begin{figure}[h!]
\includegraphics[clip,width=0.48\textwidth,angle=0]{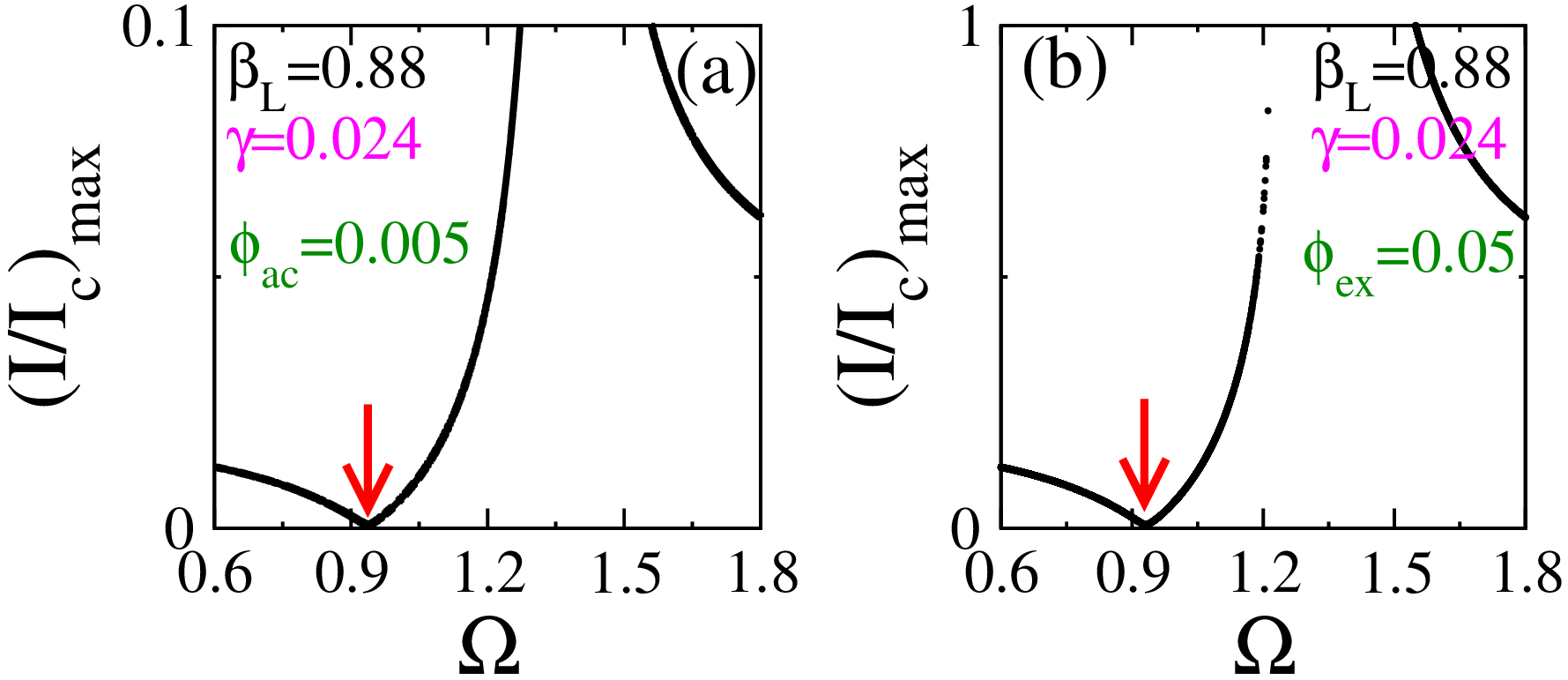}
\caption{(Color online)
Enlargement of Figs. \ref{fig999}(a) and (b) around the region of the anti-resonance.
Same parameters as in Fig. \ref{fig999}.
\label{fig998} } 
\end{figure}

The corresponding single SQUID equation is obtained from Eq. (\ref{06}) for $\lambda_0=0$,
so that ${\bf \hat{\Lambda}} ={\bf \hat{\Lambda}}^{-1} ={\bf I_N}$, where ${\bf I_N}$ is 
the $N\times N$ unit matrix, and by dropping the subscript $n$, as 
\begin{eqnarray}
\label{99}
  \ddot{\phi} +\gamma \dot{\phi} +\phi +\beta \sin\left( 2\pi \phi \right) 
    =\phi_{dc}+\phi_{ac} \cos(\Omega \tau ) .
\end{eqnarray}
The current $I$ in the SQUID ring is given by the flux-balance condition \cite{Lazarides2013b}
\begin{eqnarray}
\label{98}
  \phi =\phi_{ext} +\beta i , 
\end{eqnarray}
where $i=I/I_c$. For fixed rescaled SQUID parameter $\beta$, the nonlinear effects become
stronger with increasing amplitude of the ac field, $\phi_{ac}$. In that case, the current 
amplitude $i_{max}$ - driving frequency $\Omega$ characteristic (resonance curve) becomes 
strongly hysteretic and several simultaneously stable branches appear. This multistability 
effect is illustrated in Fig. \ref{fig999}, in which two such characteristics are shown in 
(a) and (b) for weak and strong nonlinearity, respectively. 
%%%%%%%%%%%%%%%%%%%%%%%%%%%%%%%%%%%%%%%%%%%%%%%%%%%%%%%%%%%%%%%%%%%%%%%%%%%%%%%%%%%%%%%%%%
%%%%%%%%%%%%%%%%%%%%%%%%%%%%%%%%%%%%%%%%%%%%%%%%%%%%%%%%%%%%%%%%%%%%%%%%%%%%%%%%%%%%%%%%%%
In Fig. \ref{fig999}(a), the SQUID is close to the linear regime, in which the resonance
curve is single-valued, in the sense that each value of $\Omega$ corresponds to only one 
value of $(I/I_c)_{max}$. Hysteretic effects are not visible, however the curve is not 
symmetric as it should be in the truly linear regime. Note that from Eq, (\ref{99}),
linearized around $\phi \simeq 0$, we get for the single SQUID resonance frequency the 
expression $\Omega_{SQ}= \sqrt{1 +\beta_L}$, which gives $\Omega_{SQ}= 1.37$ for the 
parameters of Fig. \ref{fig999}. In Fig. \ref{fig999}(b), in which the ac field amplitude
has been increased by an order of magnitude, strongly nonlinear effects become readily 
apparent. At least five stable branches can be identified in a narrow frequency region 
around unity, i.e., around the geometrical (inductive-capacitive, $LC$) resonance frequency.
The upper branches, which are extremely sensitive to perturbations, correspond to very
high values of the current amplitude, which leads the Josephson junction of the SQUID to
its normal state. Although the``resonance'' region in the strongly nonlinear regime has 
been shifted to the left, as compared with that in the weakly, almost linear regime,
the location of the anti-resonance has not been changed. This can be observed clearly in
Fig. \ref{fig998}, where the anti-resonance frequency region has been enlarged in both
regimes. The current amplitude anti-resonance makes itself apparent as a well defined dip
in the characteristics, with a minimum that almost reaches zero. The effect of 
anti-resonance has been observed in nonlinearly coupled oscillators subjected to a periodic 
driving force \cite{Woafo1998} and in parametrically driven nonlinear oscillators 
\cite{Chakraborty2013}. However, it has never before been observed in a single, periodically 
driven nonlinear oscillator. The knowledge of the location of anti-resonance(s) as well as 
the resonance(s) of an oscillator or a system of oscillators, beyond their theoretical 
interest, they are of importance in device applications.
Certainly they have significant implications for the SQUID metamaterials which properties 
are determined by those of their elements (i.e., the SQUIDs). When the SQUIDs are in an 
anti-resonant state, in which the induced current is zero, they do not absorb energy from 
the applied field which can transpass the metamaterial almost unaffected. Thus, in such 
a state, the SQUID metamaterial is transparent to the applied magnetic flux as has been 
already observed experimentally \cite{Zhang2015}. Moreover, since the anti-resonance 
frequency is not affected by the amplitude of the ac field, the transparency can be 
observed for a wide range of applied rf flux. In the strongly nonlinear regime, the 
anti-resonance lies in the multistability region (see Fig. \ref{fig999}b), due to the frequency 
shift of the SQUID resonance because of the strong nonlinearity.
In that case, the transparency of the metamaterial may be turned on and off as it has 
been already discussed in Ref. \cite{Zhang2015}. Thus, the concept of the anti-resonance 
serves for making a connection between an important SQUID metamaterial property and a 
fundamental nonlinear dynamical property of oscillators.        
%%%%%%%%%%%%%%%%%%%%%%%%%%%%%%%%%%%%%%%%%%%%%%%%%%%%%%%%%%%%%%%%%%%%%%%%%%%%%%%%%%%%%%%%%%
%%%%%%%%%%%%%%%%%%%%%%%%%%%%%%%%%%%%%%%%%%%%%%%%%%%%%%%%%%%%%%%%%%%%%%%%%%%%%%%%%%%%%%%%%%

%%%%%%%%%%%%%%%%%%%%%%%%%%%%%%%%%%%%%%%%%%%%%%%%%%%%%%%%%%%%%%%%%%%%%%%%%%%%%%%%%%%%%%%%%%
For the time integration of Eq. (\ref{99}) of the single SQUID, a fourth-order Runge-Kutta
algorithm with constant time-step has been used. The same algorithm is also used for the 
simulation of the SQUID metamaterial which is modeled by Eq. (\ref{06}). The latter are  
implemented with the following boundary conditions
\begin{equation}
\label{11} 
   \phi_0 (\tau) =0 , \qquad \phi_{N+1} (\tau) =0 ,
\end{equation}
to account for the termination of the structure in a finite system. 
The parameter values used in the simulations of the single SQUID and the SQUID 
metamaterial are in the region in which the relevant experiments were carried out
\cite{Trepanier2013,Zhang2015}. 
In Ref. \cite{Trepanier2013}, the values of the SQUIDs' inductance, resistance, and 
capacitance are $L=0.12~nH$, $R=840~\Omega$, and $C=0.65~pF$, respectively, while the 
critical current is $3.7~\mu A$ and $1.2~\mu A$ at temperature $T=4.2~K$ and $6.5~K$, 
respectively. Inserting these values in the first of Eqs. (\ref{14}), we get for $\beta_L$ 
the values $0.44$ and $1.3$ at $T=4.2~K$ and $6.5~K$, respectively. For the SQUID 
metamaterial we use 
$\beta_L =0.7$, which lies in between those values, while our value of the amplitude of 
the alternating field $\phi_{ac}=0.015$ lies within the range of values used in the 
experiments ($\phi_{ac} \simeq 0.006 - 0.05$). The coupling coefficient between nearest 
neighbors in Ref. \cite{Trepanier2013} has been estimated to be $\lambda_0 \simeq-0.02$, 
using a simple approximation scheme in which the SQUIDs are regarded as thin rings. 
However, a large part of the area of the actual SQUID metamaterial is filled by 
superconducting material that the field cannot penetrate; it is thus expected that more 
magnetic flux than that predicted by the simple approximation will find its way through 
the SQUID rings, increasing thus considerably $\lambda_0$ (we use $\lambda_0=-0.05$). 
The value of the loss coefficient in our simulations for the SQUID metamaterial is 
$\gamma \simeq 0.002$ (for the single SQUID $\gamma \simeq 0.024$), while 
from the second of Eqs. (\ref{14}) and the values given in Ref. \cite{Trepanier2013} we get
$\sim 0.02$. Although the losses in the SQUIDs can be reduced considerably by lowering the
temperature without affecting much the critical currents of the junctions, we have checked 
that chimera states also exist for $\gamma$ of the order of $\sim 0.02$ when compensated
by a larger ac field amplitude $\phi_{ac}$.
Note that the parameters used in the simulations of the one-dimensional SQUID metamaterial 
are compared with those in Ref. \cite{Trepanier2013}, in which the experiments have been 
carried out on two-dimensional arrays, merely to show that these are realistic. In this work, 
we do not attempt to simulate a particular SQUID metamaterial. Moreover, the dimensionality 
of the system does not affect significantly the estimation of the parameters that are 
necessary for the simulations, since they can be estimated either by the individual SQUID 
properties (i.e., $\beta$, $\gamma$) or by a pair of SQUIDs (i.e., $\lambda_0$).

%%%%%%%%%%%%%%%%%%%%%%%%%%%%%%%%%%%%%%%%%%%%%%%%%%%%%%%%%%%%%%%%%%%%%%%%%%%%%%%%%%%%%%%%%%
The spatial coherence and incoherence of the chimera state can be characterized by a 
real-valued local order parameter \cite{Omelchenko2011}
\begin{equation}
\label{eq:locord}
   Z_n=\frac{1}{2\delta} \sum_{|m-n| \le \delta} e^{i2\pi \phi_m} , 
       \quad n=\delta+1,\dots, N-\delta .
\end{equation}
We use a spatial average with a window size of $\delta=10$ elements. A local order parameter 
$Z_n=1$ indicates that the $n_{th}$ unit belongs to the coherent part of the chimera state, 
while $Z_n$ is less than 1 for incoherent parts. Note that in our system we have no periodic 
boundary conditions like in \cite{Omelchenko2011}; therefore Eq.~\ref{eq:locord} holds for 
the SQUIDs with indices running from $n=\delta+1,\dots, N-\delta$. For the SQUIDs close to 
boundaries 
of the structure, the calculation of the local order parameter is modified as follows:
$Z_n=\frac{1}{\delta} \sum\limits_{m=n}^{n+\delta} e^{i2\pi \phi_m}$ for $n=1,\dots,\delta$ and
$Z_n=\frac{1}{\delta} \sum\limits_{m=n-\delta}^{n} e^{i2\pi \phi_m}$ for $n=N-\delta+1,\dots,N$.
A measure for the synchronization level in the whole metamaterial is given by magnitude of
the global Kuramoto order parameter $|\Psi(\tau)|$, where $\Psi(\tau)$ is defined as
\begin{equation}
\label{eq:globord} 
 \Psi(\tau) = \frac{1}{N} \sum_{m=1}^N e^{i [2\pi \phi_m (\tau)]} .
\end{equation}
This measure will be used in Sec.~\ref{sec:sec4} where the transition from global 
synchronization to desynchronization will be discussed.
%%%%%%%%%%%%%%%%%%%%%%%%%%%%%%%%%%%%%%%%%%%%%%%%%%%%%%%%%%%%%%%%%%%%%%%%%%%%%%%%%%%%%%%%%%
\begin{figure}[t!]
\includegraphics[clip,width=0.31\textwidth,angle=-90]{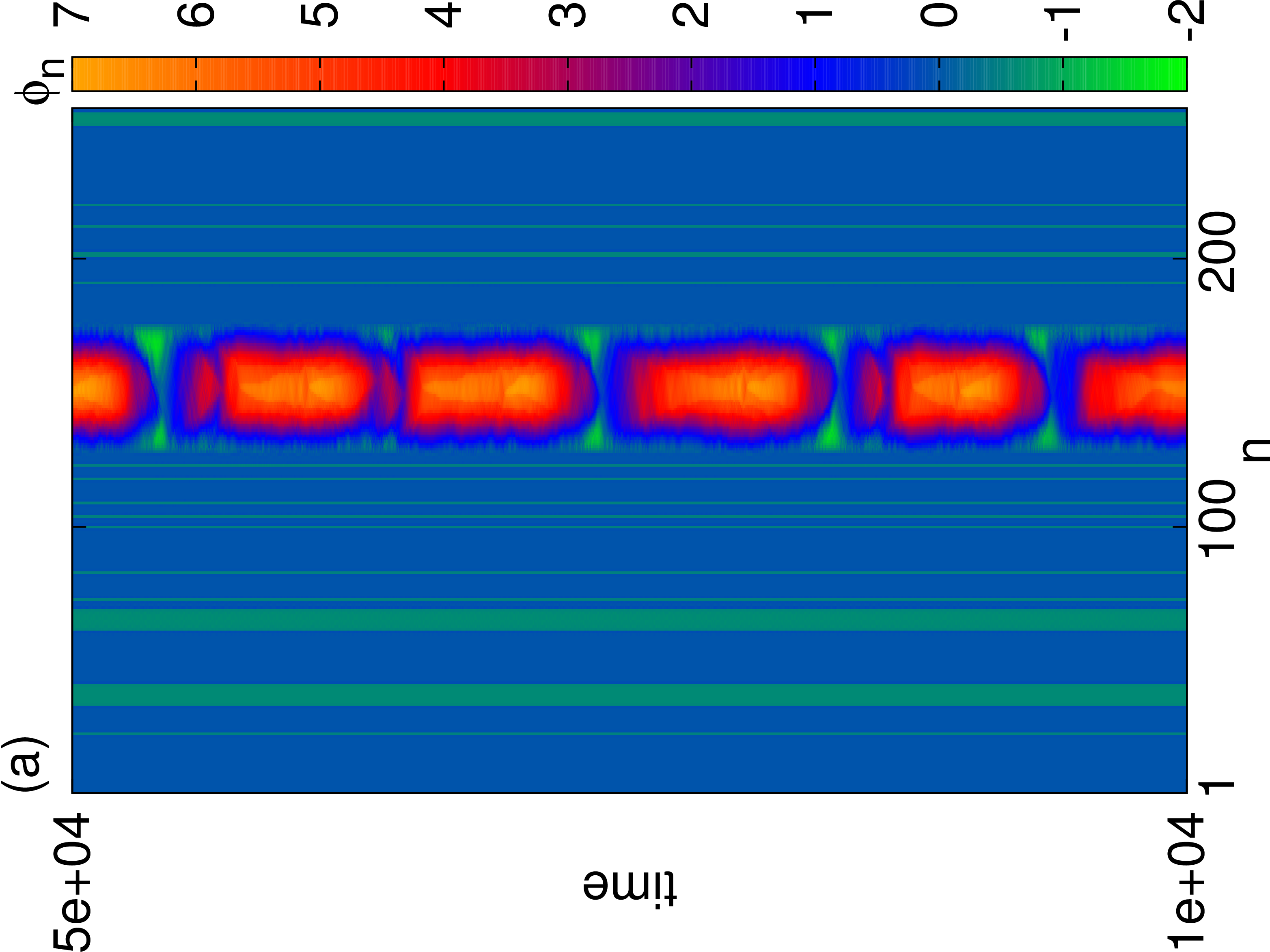}
\includegraphics[clip,width=0.31\textwidth,angle=-90]{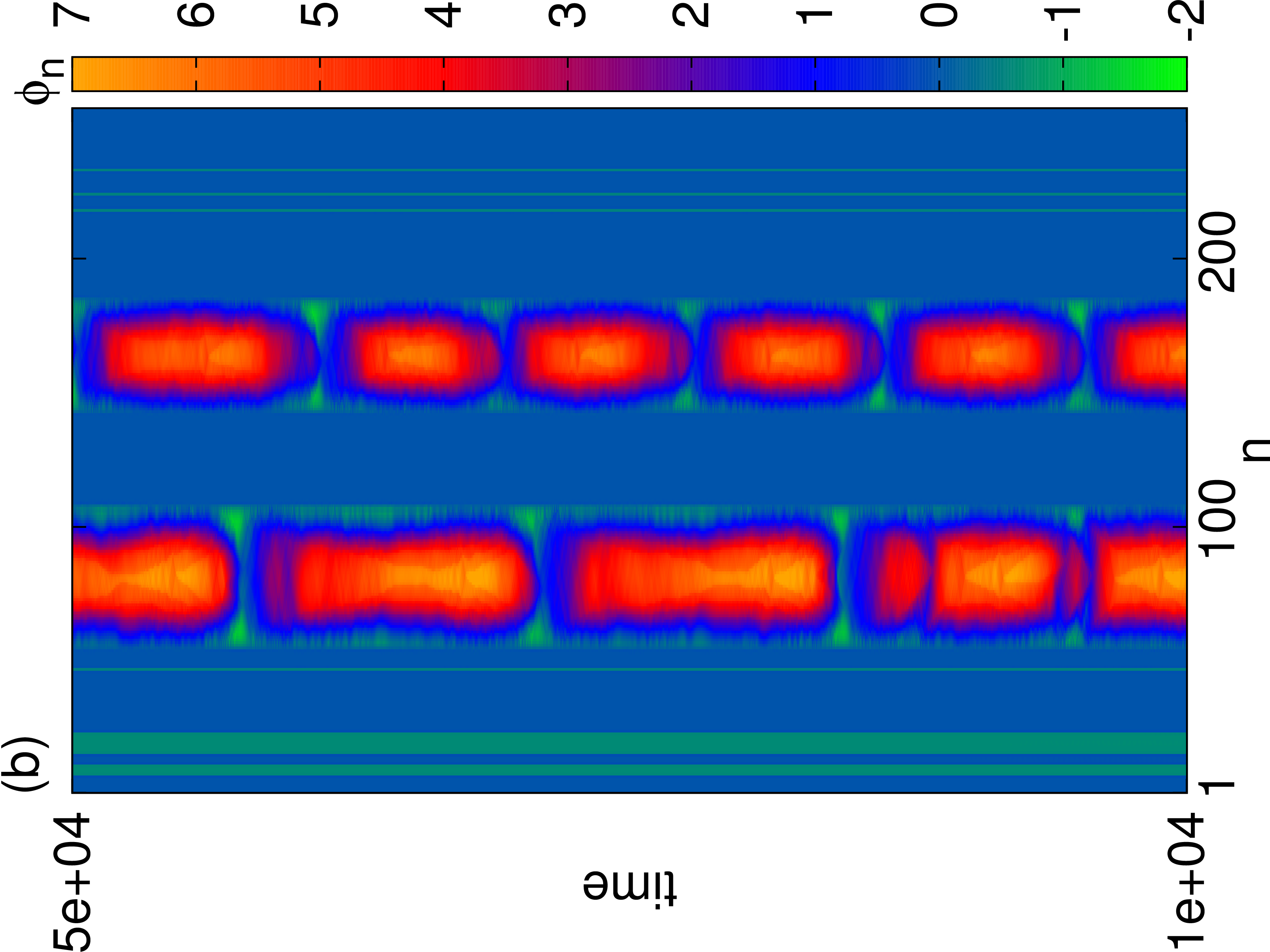}\\
\includegraphics[clip,width=0.31\textwidth,angle=-90]{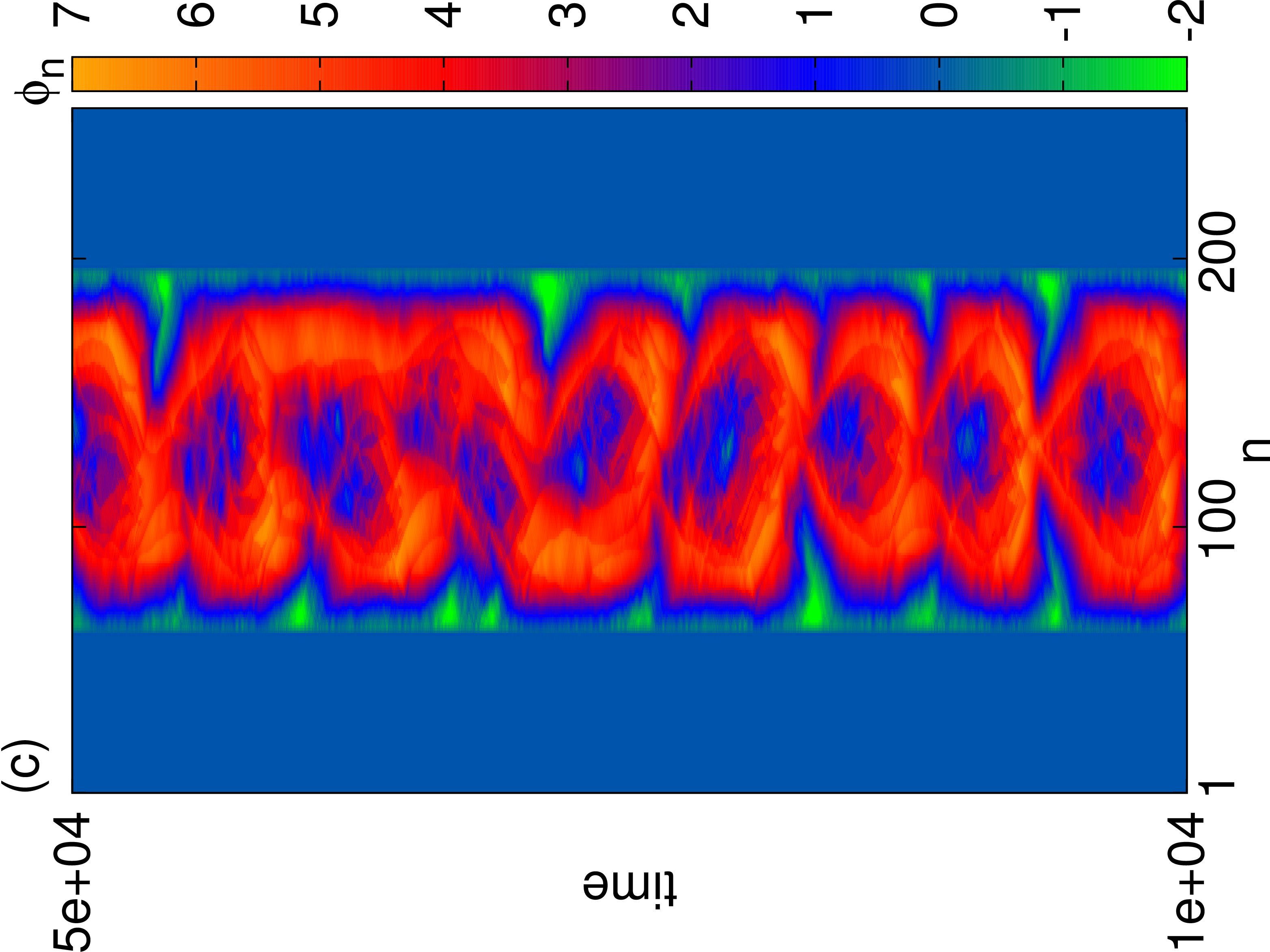}
\includegraphics[clip,width=0.31\textwidth,angle=-90]{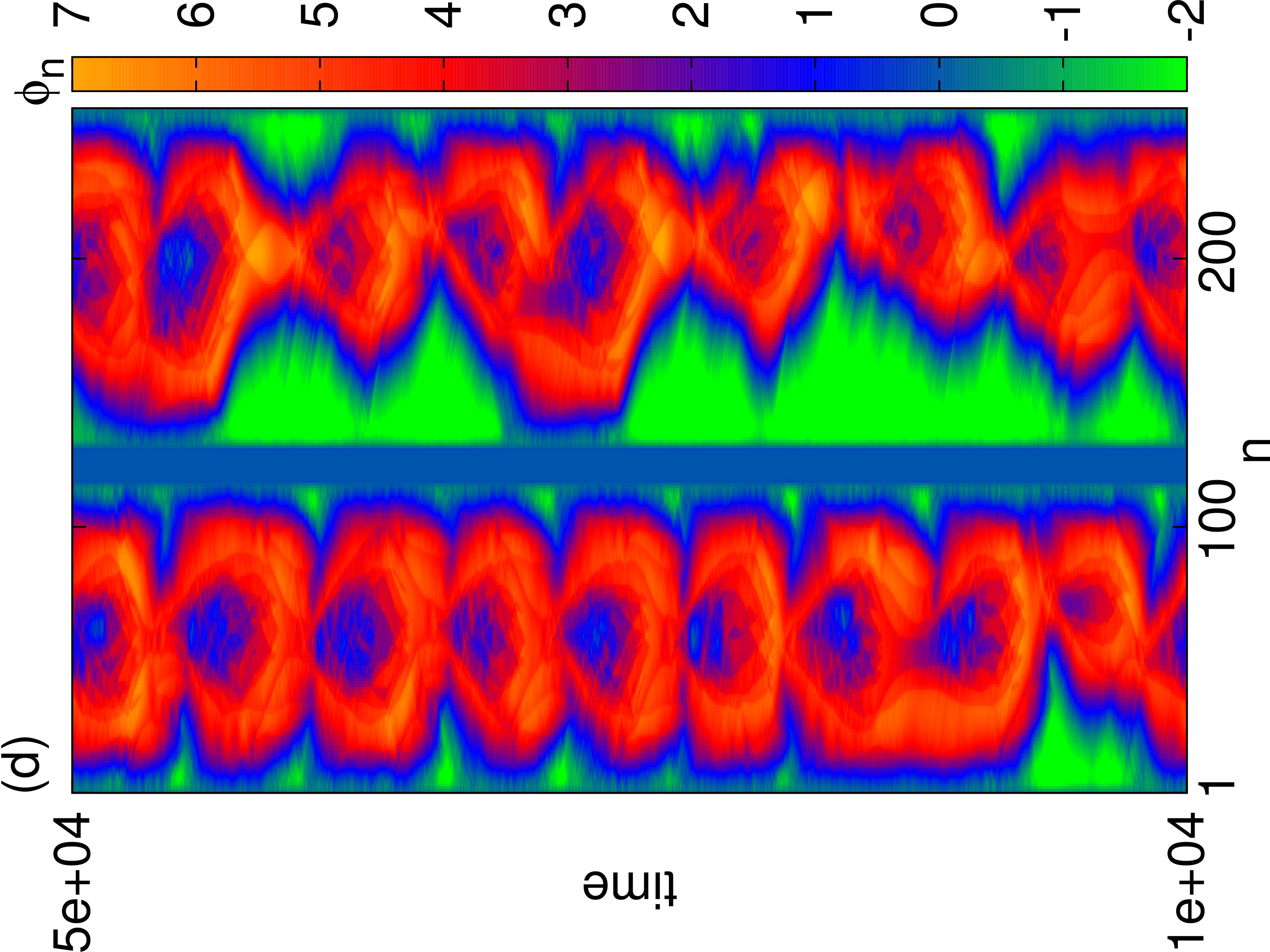}\\
\includegraphics[clip,width=0.31\textwidth,angle=-90]{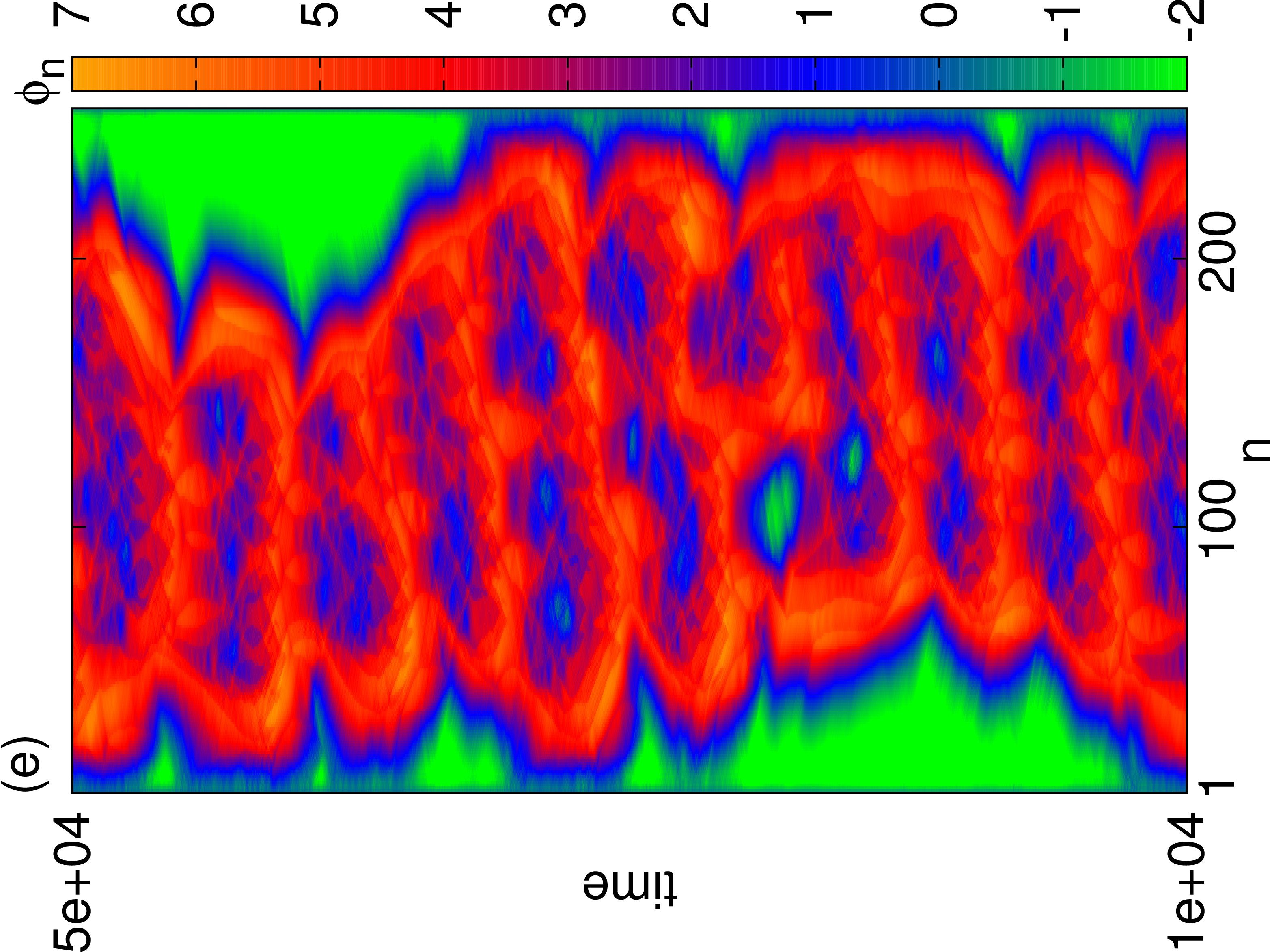}
\includegraphics[clip,width=0.31\textwidth,angle=-90]{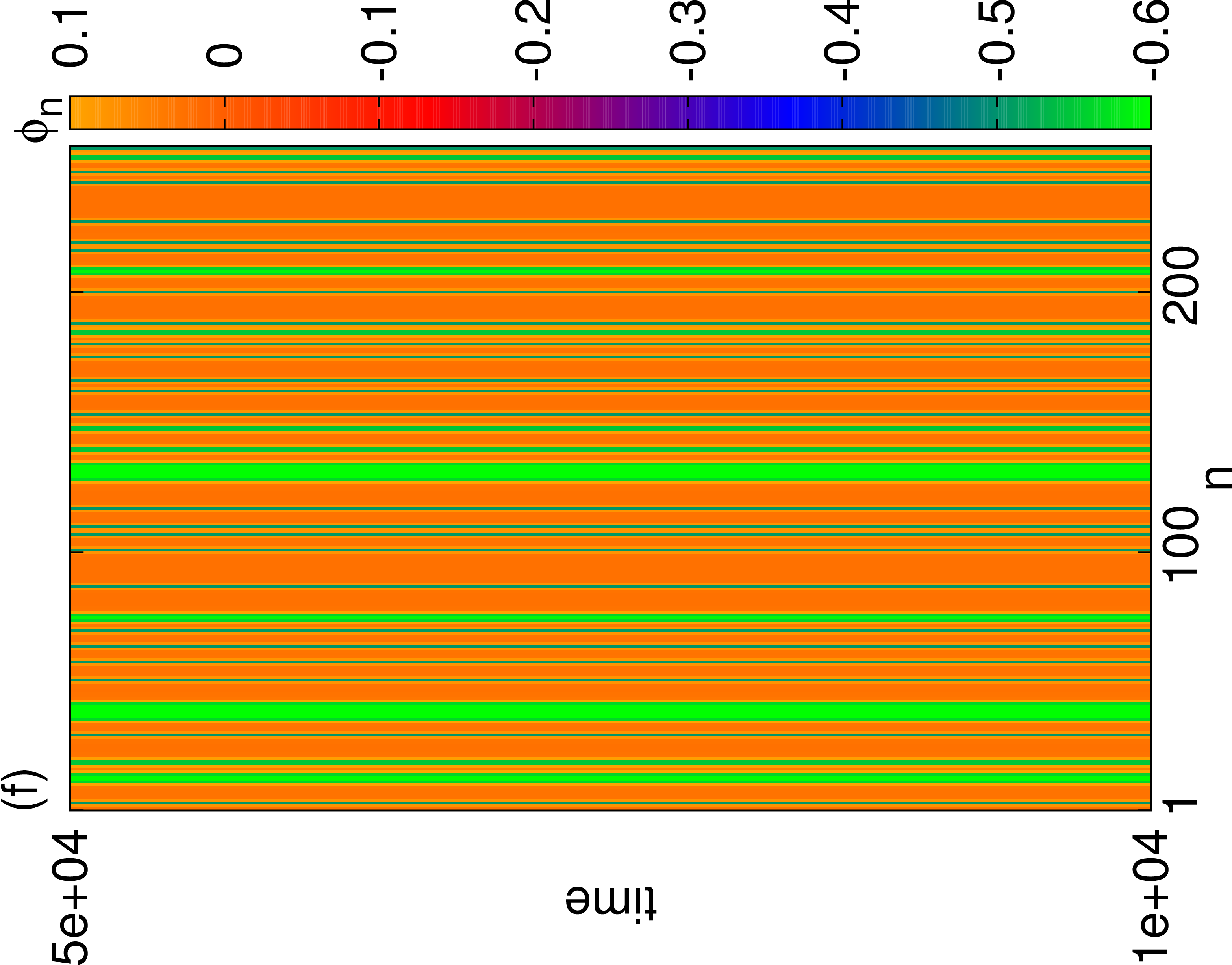}
\caption{(Color online)
Space-time plots for the flux $\phi_n$ over the whole SQUID metamaterial for different 
initial conditions. (a), (c) States with one incoherent region, (b), (d) States with two 
incoherent regions, (e) Travelling incoherent state, (f) Solitary state.
Parameter values are: $T=5.9$, $N=256$, $\gamma=0.0021$, $\lambda_0=-0.05$, $\beta = 0.1114$ 
($\beta_L\simeq 0.7$), $\phi_{ac}=0.015$, $\phi_{dc}=0.0$.
\label{fig2} }
\end{figure}
%%%%%%%%%%%%%%%%%%%%%%%%%%%%%%%%%%%%%%%%%%%%%%%%%%%%%%%%%%%%%%%%%%%%%%%%%%%%%%%%%%%%%%%%%%

%%%%%%%%%%%%%%%%%%%%%%%%%%%%%%%%%%%%%%%%%%%%%%%%%%%%%%%%%%%%%%%%%%%%%%%%%%%%%%%%%%%%%%%%%%
\begin{figure}[t!]
\includegraphics[clip,width=0.45\linewidth,angle=0]{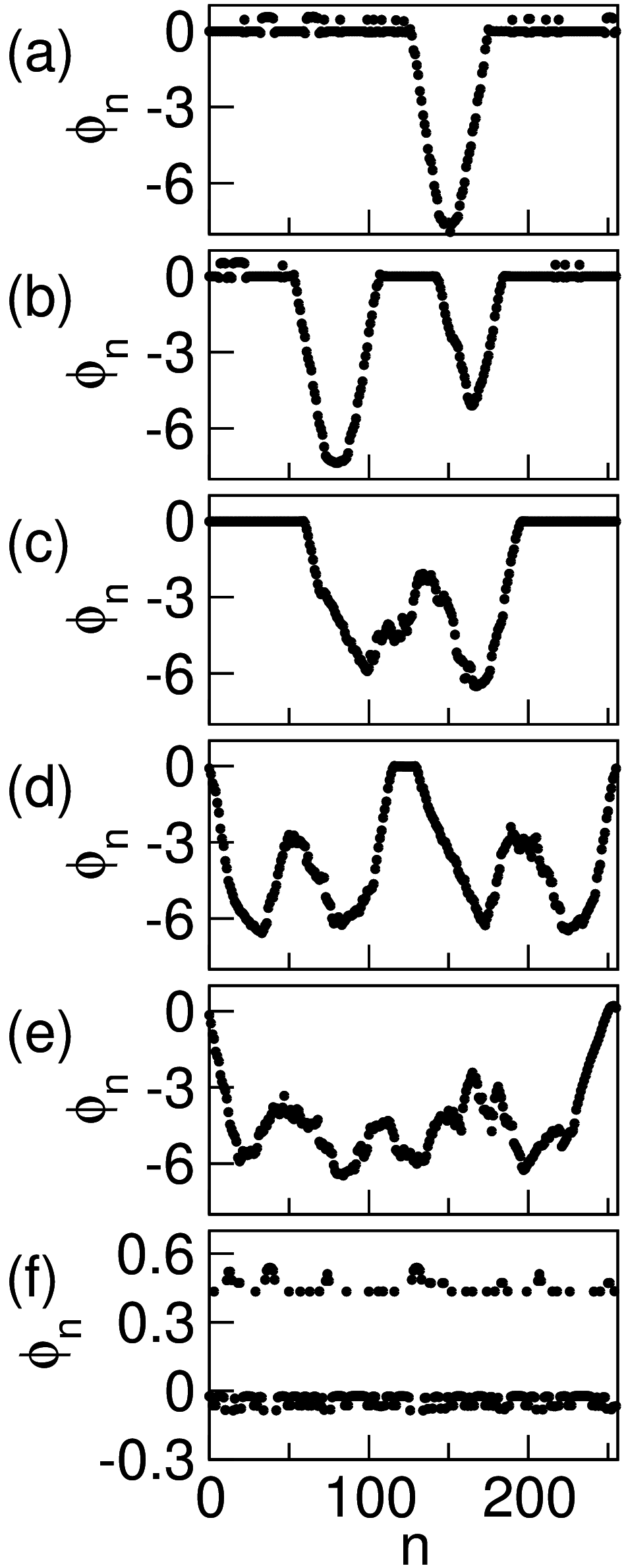}
\hspace{0.4cm}
\includegraphics[clip,width=0.45\linewidth,angle=0]{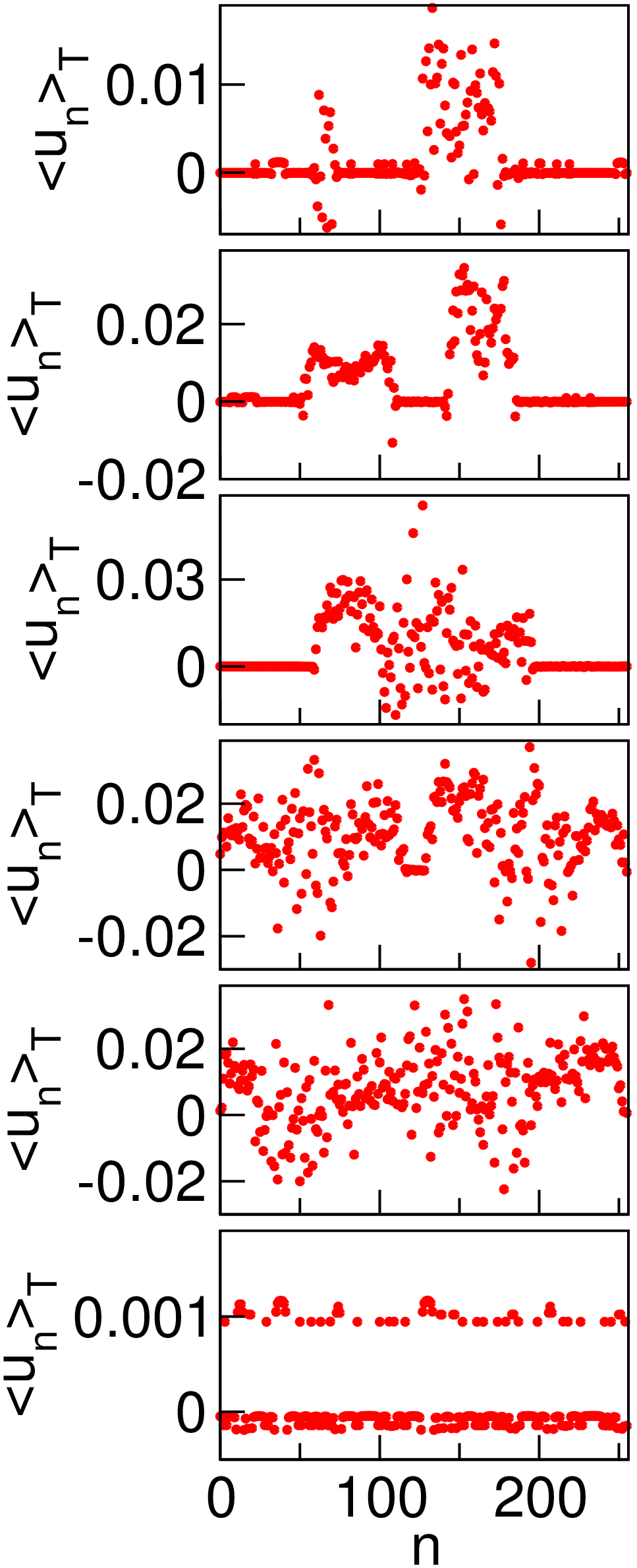}
\caption{(Color online) 
Snapshots of the spatial profile of the fluxes $\phi_n$ corresponding to Fig.~\ref{fig2} 
(left panel) and derivative of the phases $\dot{\phi}_n$ averaged over one period of the 
driving force $T$. Parameters as in Fig.~\ref{fig2}.
\label{fig3}}
\end{figure}
%%%%%%%%%%%%%%%%%%%%%%%%%%%%%%%%%%%%%%%%%%%%%%%%%%%%%%%%%%%%%%%%%%%%%%%%%%%%%%%%%%%%%%%%%%

%%%%%%%%%%%%%%%%%%%%%%%%%%%%%%%%%%%%%%%%%%%%%%%%%%%%%%%%%%%%%%%%%%%%%%%%%%%%%%%%%%%%%%%%%%
\section{Multistability and Synchronization}
\label{sec:sec3}
The individual SQUID is a highly nonlinear oscillator exhibiting multistability in a 
certain parameter regime. This is very crucial for the occurrence of the chimera states
when considering the coupled system under investigation. The number of possible states in 
a SQUID metamaterial is not merely the sum of the combinations of individual SQUID states, 
since their collective behavior may provide many more possibilities.
Depending on the choice of initial conditions, various space-time flux patterns may be 
obtained, shown in Fig.~\ref{fig2}, where the evolution of the $\phi_n$s is monitored 
at times equal to multiples of one driving period $T=2\pi/\Omega$. In particular, 
Fig.~\ref{fig2}(a) corresponds to a typical pattern with two distinct domains: 
a cluster of SQUIDS located around $n=150$ in which the fluxes oscillating with high 
amplitude coexists with the rest of the array which oscillates with low-flux amplitude. 
The latter part of the array is not completely homogeneous since small clusters and 
individual SQUIDs perform slightly higher amplitude oscillations, which manifest themselves 
through the thin green lines. In Fig.~\ref{fig2}(b) a similar pattern with two clusters 
of high amplitude oscillations is depicted. Single and double-headed chimera states with 
larger sizes of incoherent clusters may also be achieved, as shown in Fig.~\ref{fig2}(c)
and (d), respectively. A coexisting traveling pattern can be seen in Fig.~\ref{fig2}(e), 
where the largest part of the metamaterial is occupied by an incoherent cluster with 
varying size and position in time. Finally Fig.~\ref{fig2}(f) demonstrates a pattern 
of low-amplitude oscillations with multiple so-called solitary states \cite{Maistrenko2014}, 
where many SQUIDs 
have escaped from the main synchronized cluster and perform oscillations of slightly 
higher amplitudes (depicted by the light green stripes in the otherwise orange background).

%%%%%%%%%%%%%%%%%%%%%%%%%%%%%%%%%%%%%%%%%%%%%%%%%%%%%%%%%%%%%%%%%%%%%%%%%%%%%%%%%%%%%%%%%%
\begin{figure}[h!]
\includegraphics[clip,width=0.3\textwidth,angle=-90]{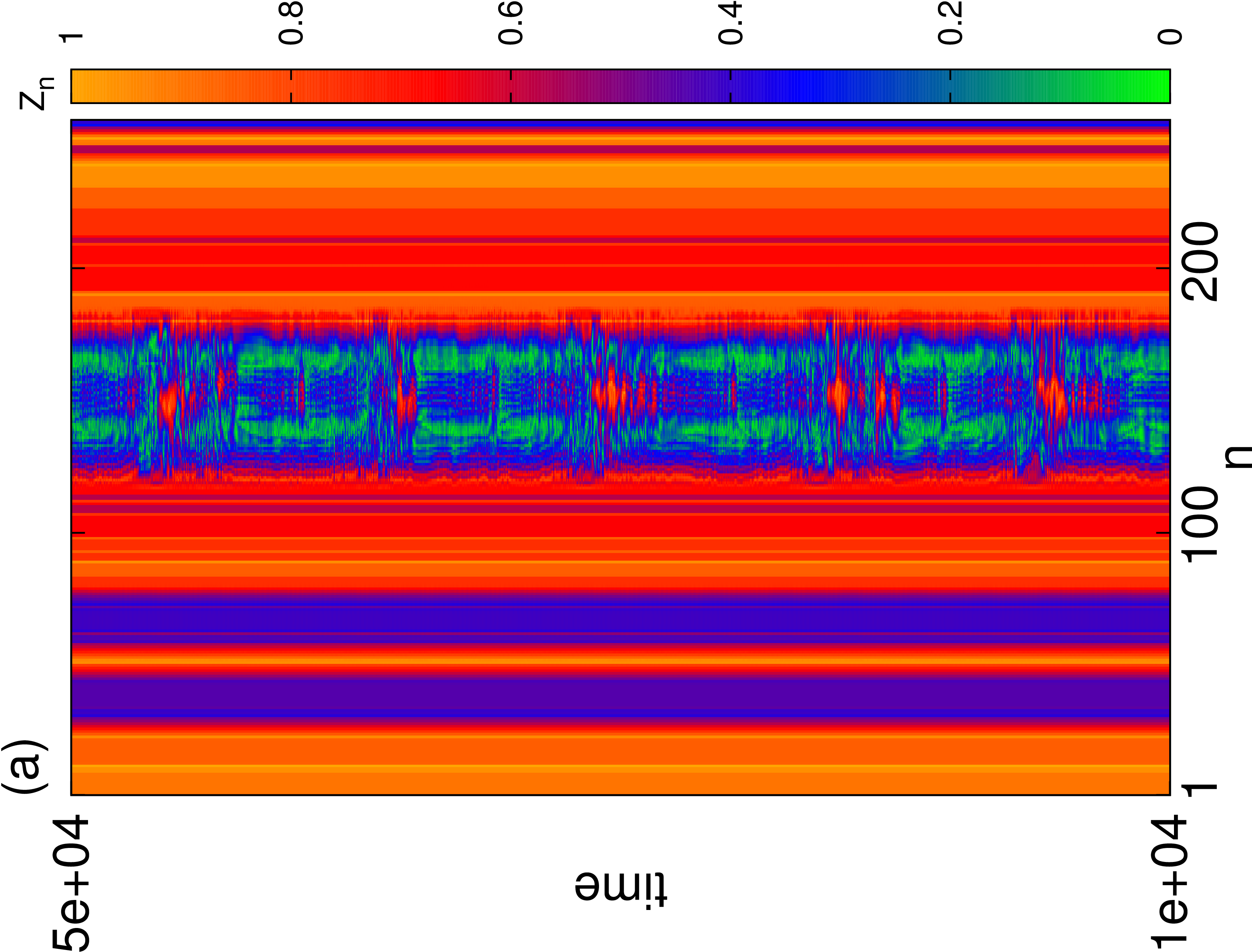}
\includegraphics[clip,width=0.3\textwidth,angle=-90]{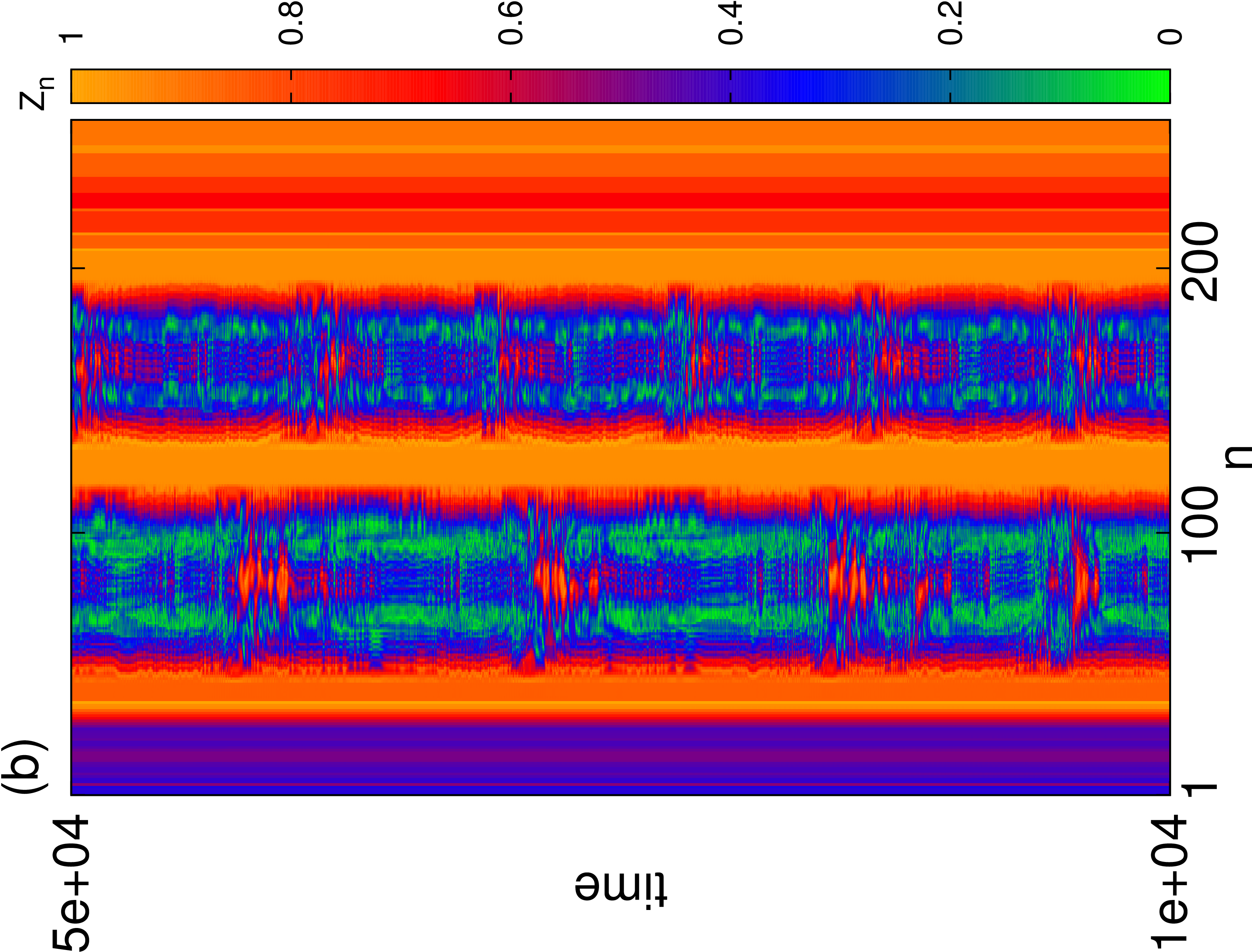}\\
\includegraphics[clip,width=0.3\textwidth,angle=-90]{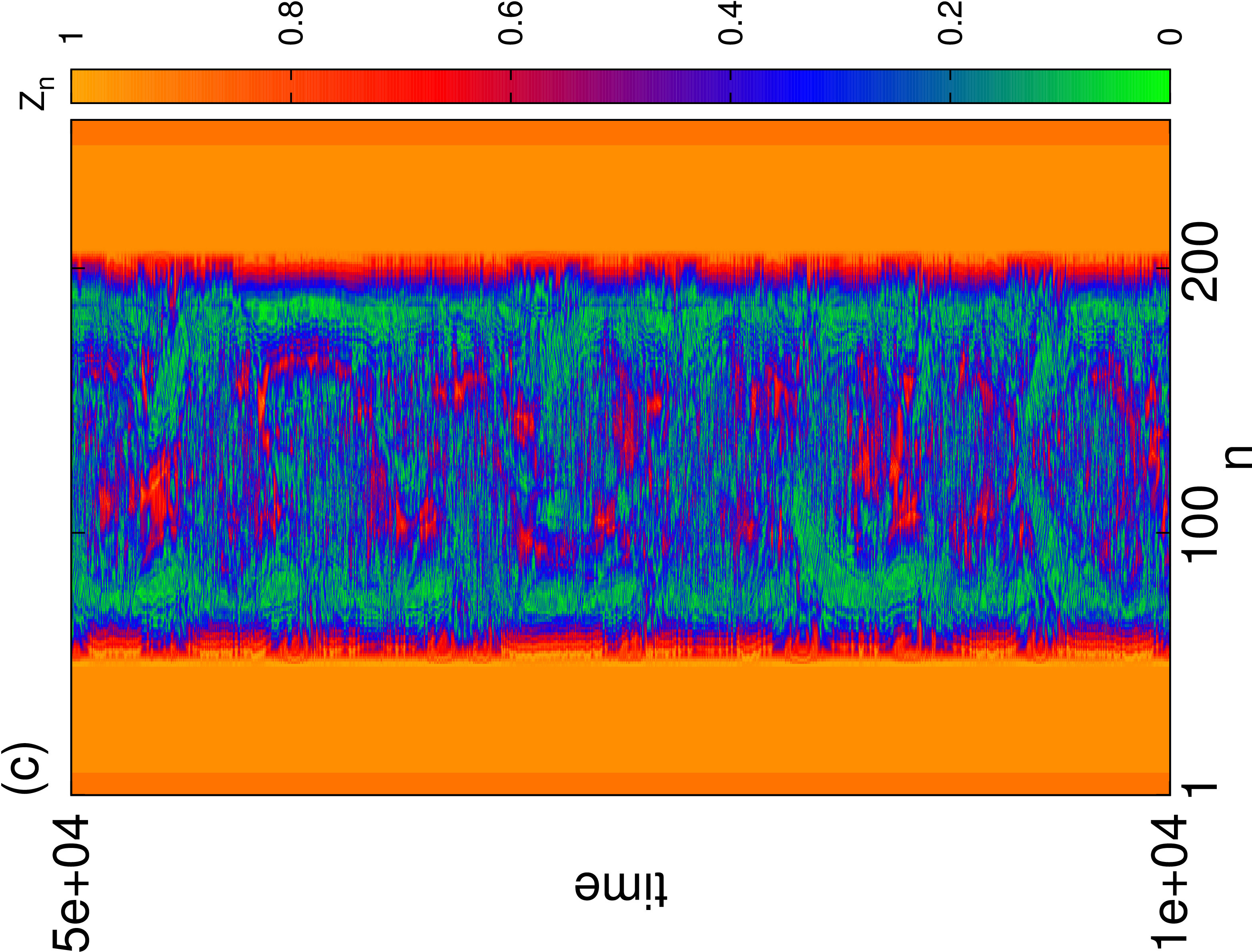}
\includegraphics[clip,width=0.3\textwidth,angle=-90]{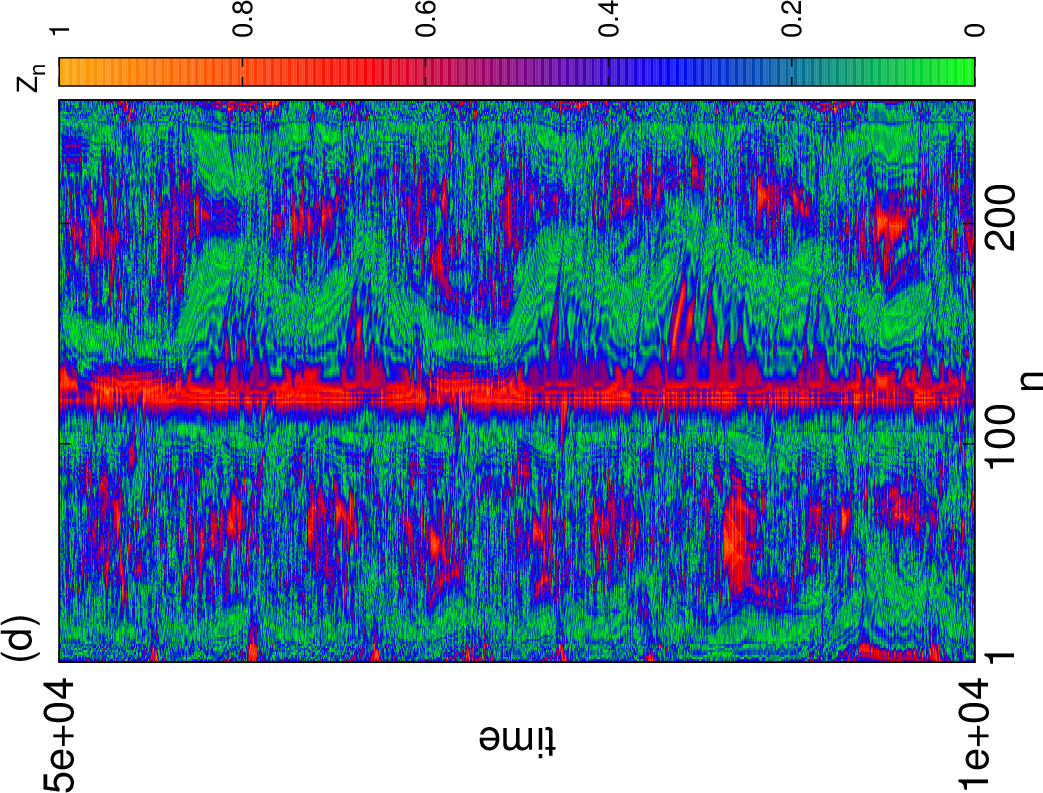}\\
\includegraphics[clip,width=0.3\textwidth,angle=-90]{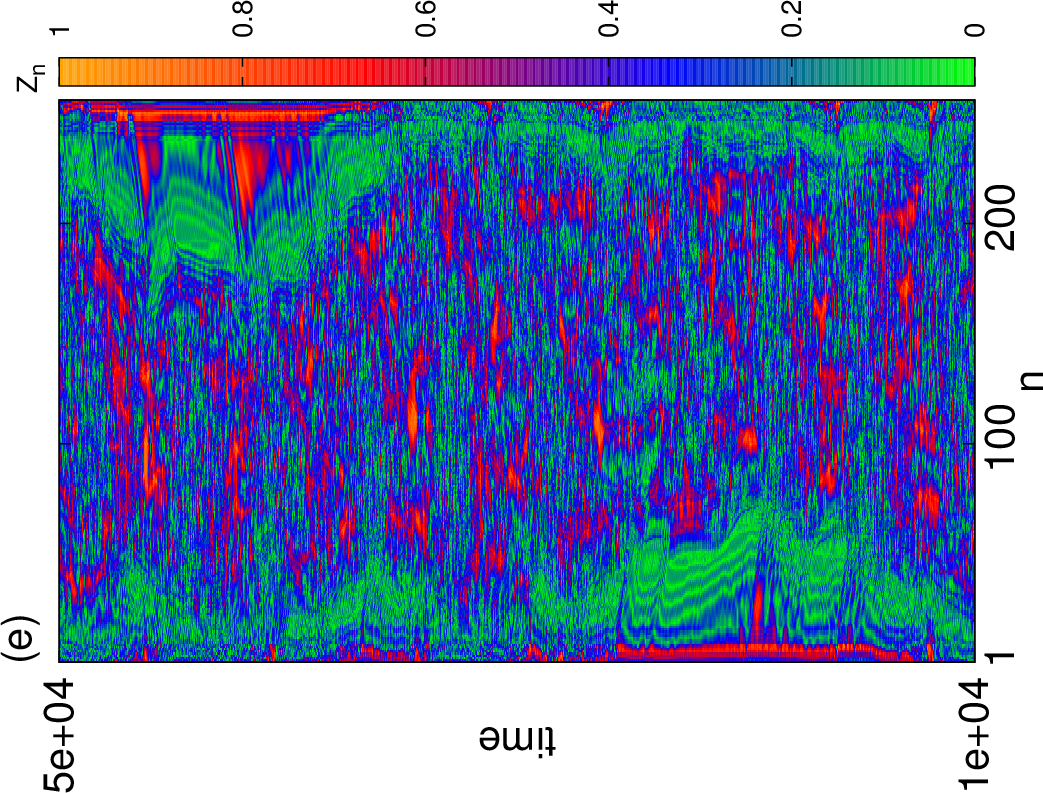}
\includegraphics[clip,width=0.3\textwidth,angle=-90]{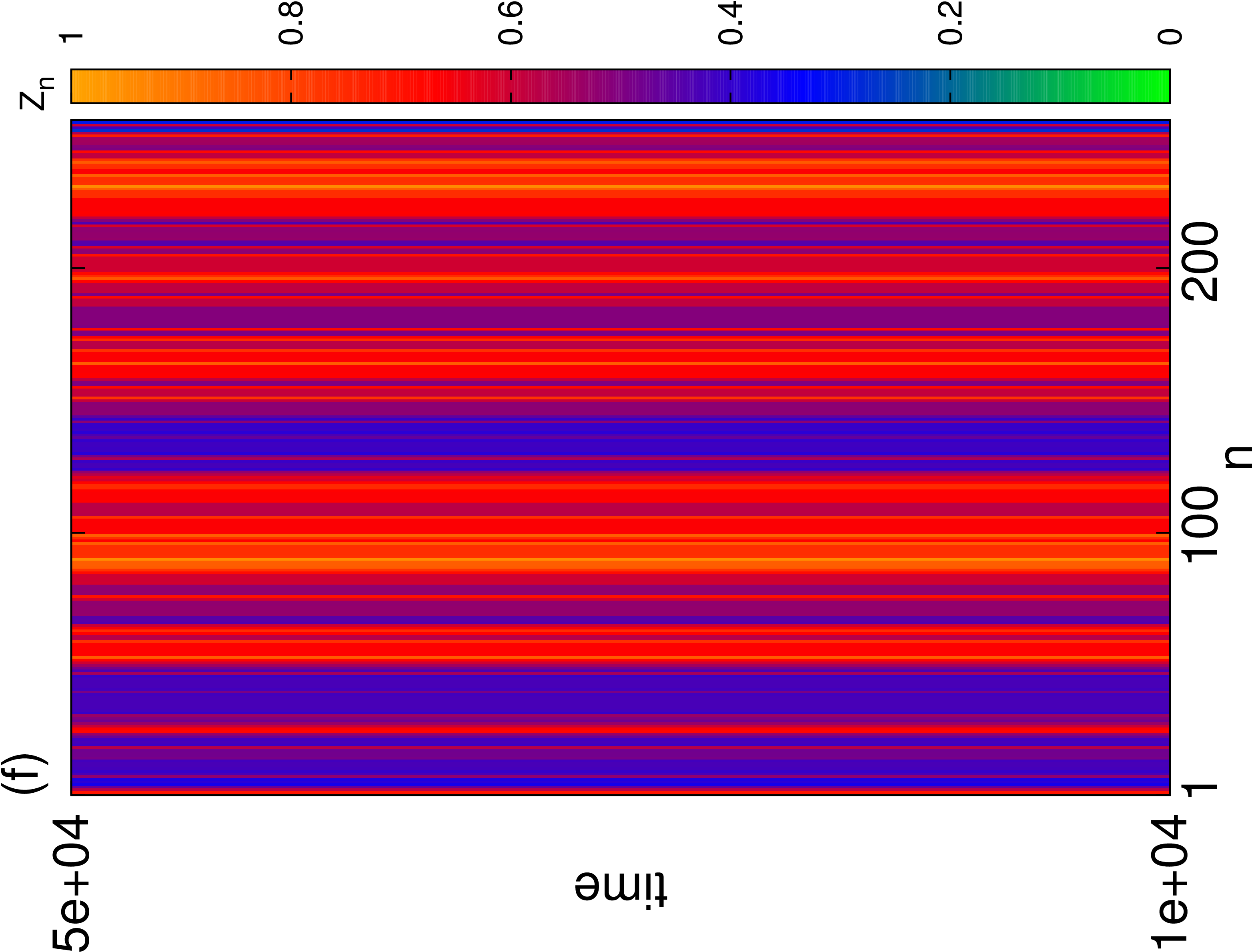}
\caption{(Color online) 
Space-time plots for magnitude of the local order parameter $|Z_n|$ of the corresponding 
states of Figure~\ref{fig2}. Parameters as in Fig.~\ref{fig2}.
\label{fig4} }
\end{figure}
%%%%%%%%%%%%%%%%%%%%%%%%%%%%%%%%%%%%%%%%%%%%%%%%%%%%%%%%%%%%%%%%%%%%%%%%%%%%%%%%%%%%%%%%%%
%%%%%%%%%%%%%%%%%%%%%%%%%%%%%%%%%%%%%%%%%%%%%%%%%%%%%%%%%%%%%%%%%%%%%%%%%%%%%%%%%%%%%%%%%%
\begin{figure}[h!]
\includegraphics[clip,width=0.48\textwidth,angle=0]{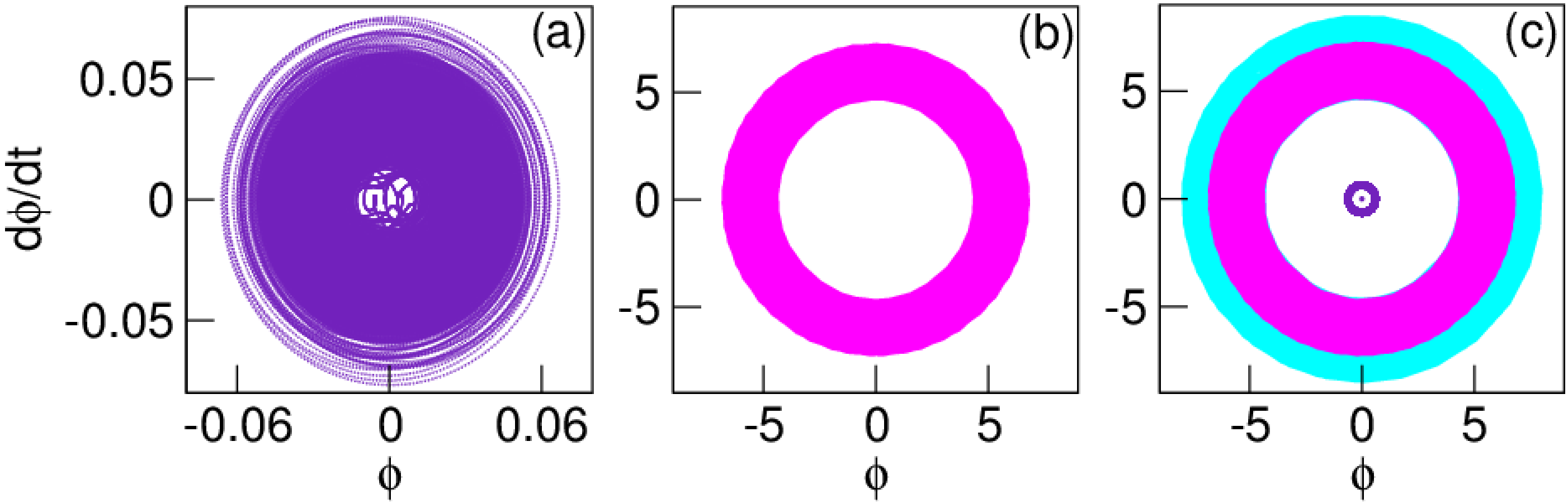}
\includegraphics[clip,width=.5\textwidth,angle=0]{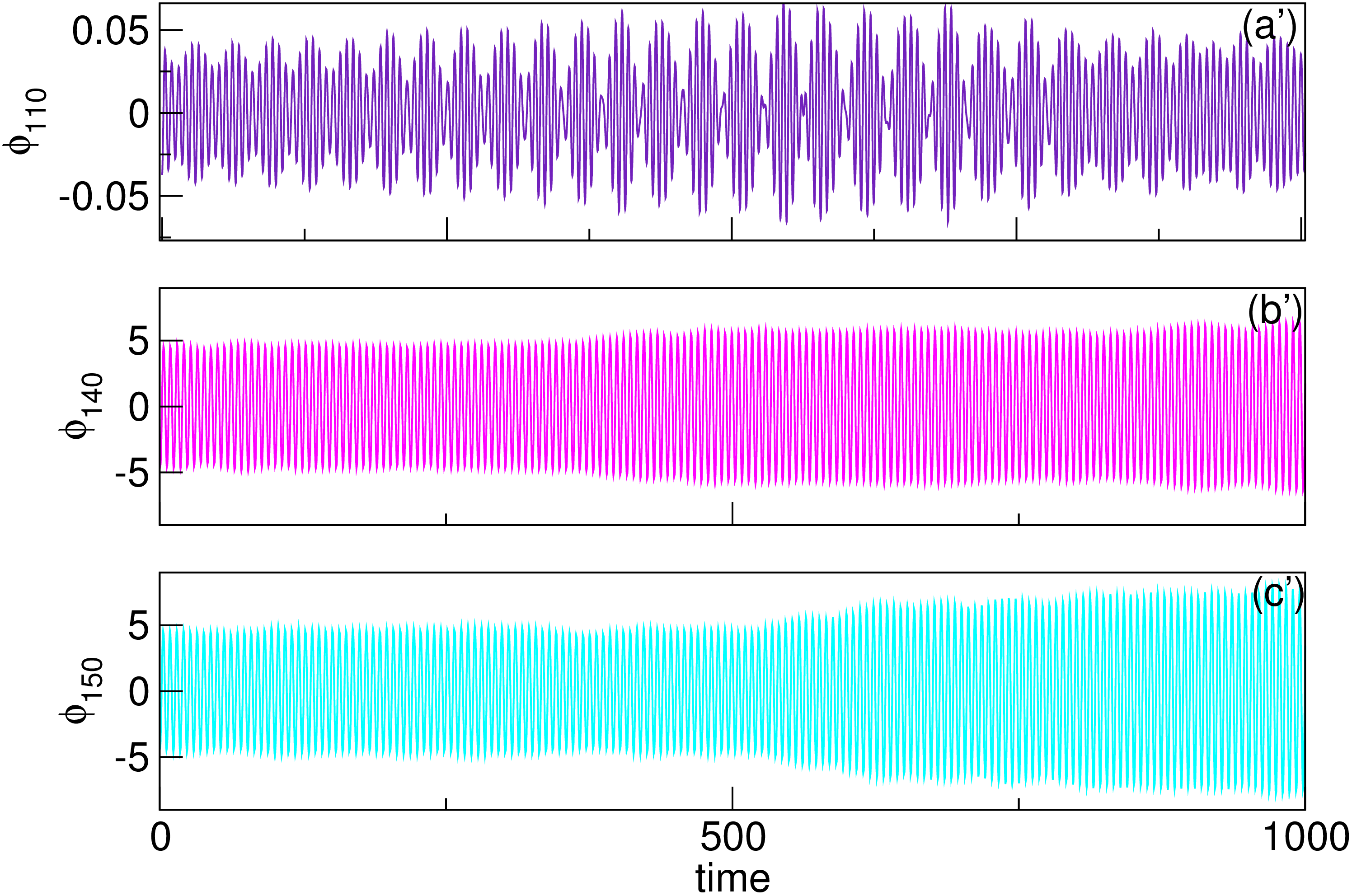}
\caption{(Color online)
Upper panel: phase portraits of SQUID oscillators at nodes $110$ (a); $140$ (b); $150$ (c). 
Lower panel: 
Corresponding time series of magnetic flux $\phi$. Parameters as in Fig.~\ref{fig2}.
\label{fig5} }
\end{figure}
%%%%%%%%%%%%%%%%%%%%%%%%%%%%%%%%%%%%%%%%%%%%%%%%%%%%%%%%%%%%%%%%%%%%%%%%%%%%%%%%%%%%%%%%%%
The corresponding snapshots of $\phi_n$ and the time-derivatives of the fluxes averaged over $T$,
$\langle \dot{\phi}_n (\tau) \rangle_T \equiv \langle v_n (\tau)\rangle_T$, at the end of 
the integration time ($\sim 5\times10^4$ time units) are shown in the 
left and right panels of Figs.~\ref{fig3}, respectively. Note that $v_n (\tau)$ is 
the instantaneous voltage across the Josephson junction of the $n-$th rf SQUID. From the 
profiles of $\langle v_n (\tau)\rangle_T$ it can be seen that the coherent nodes have 
values very close to zero, which means that the SQUIDS in the coherent part are dominated 
by the driving force, while the incoherent nodes exhibit small deviations around the 
external driving frequency.
The synchronization within the aforementioned states can be visualized through the 
space-time plots of the local order parameter (Eq.~\ref{eq:locord}), shown in 
Fig.~\ref{fig4}. Red-orange colors denote the coherent regions and blue-green colors 
the incoherent ones. These plots reveal the complexity of the synchronization levels in 
the SQUID metamaterial: For example in Fig.~\ref{fig4}(a) it can be seen that the 
incoherent region located in the center of the system {\em periodically} achieves high 
values of 
synchronization demonstrated by the orange ``islands'' within the cluster. This is related 
to metastability, which was extensively studied in \cite{LAZ15}. In the coherent cluster, 
on the other hand, we observe blue stripes of low synchronization which are evident for 
solitary states that have escaped. Note that {\em periodic synchronization}, characterized 
by periodic variation of the order parameter, has been previously observed in phase oscillator
models with external periodic driving both with and without an inertial term 
\cite{Choi1994,Hong1999}.

%%%%%%%%%%%%%%%%%%%%%%%%%%%%%%%%%%%%%%%%%%%%%%%%%%%%%%%%%%%%%%%%%%%%%%%%%%%%%%%%%%%%%%%%%%
As already pointed out, a very crucial element for the formation of chimera states in our 
system is the multistability of periodic solutions in the single SQUID. For a certain set 
of parameter values, the single SQUID may reach either a high or a low flux amplitude 
state, from basins of attraction typically smaller in the first case. Nonlocal coupling of 
SQUIDS can lead to stabilization of high flux amplitude states coexisting with low flux 
amplitude ones. Typical phase portraits and the corresponding time series for such periodic 
solutions are shown in Fig.~\ref{fig5} for three SQUID oscillators of Fig.~\ref{fig2}(a).
Figure~\ref{fig5}(a) corresponds to the SQUID at node $10$ which belongs to the 
coherent cluster of the chimera states of Fig.~\ref{fig2}(a), while Figs.~\ref{fig2}(b) 
and (c) refer to nodes on the incoherent cluster, showing high amplitude oscillations. 
The corresponding time series in Fig.~\ref{fig5}(a')-(c') show highly modulated 
quasiperiodic behavior. Additional investigations have shown evidence of chaotic dynamics 
which requires further study beyond the scope of this paper.

%%%%%%%%%%%%%%%%%%%%%%%%%%%%%%%%%%%%%%%%%%%%%%%%%%%%%%%%%%%%%%%%%%%%%%%%%%%%%%%%%%%%%%%%%%
\section{Synchronization-desynchronization transition}
\label{sec:sec4}
\begin{figure*}
\includegraphics[clip,width=0.28\textwidth,angle=-90]{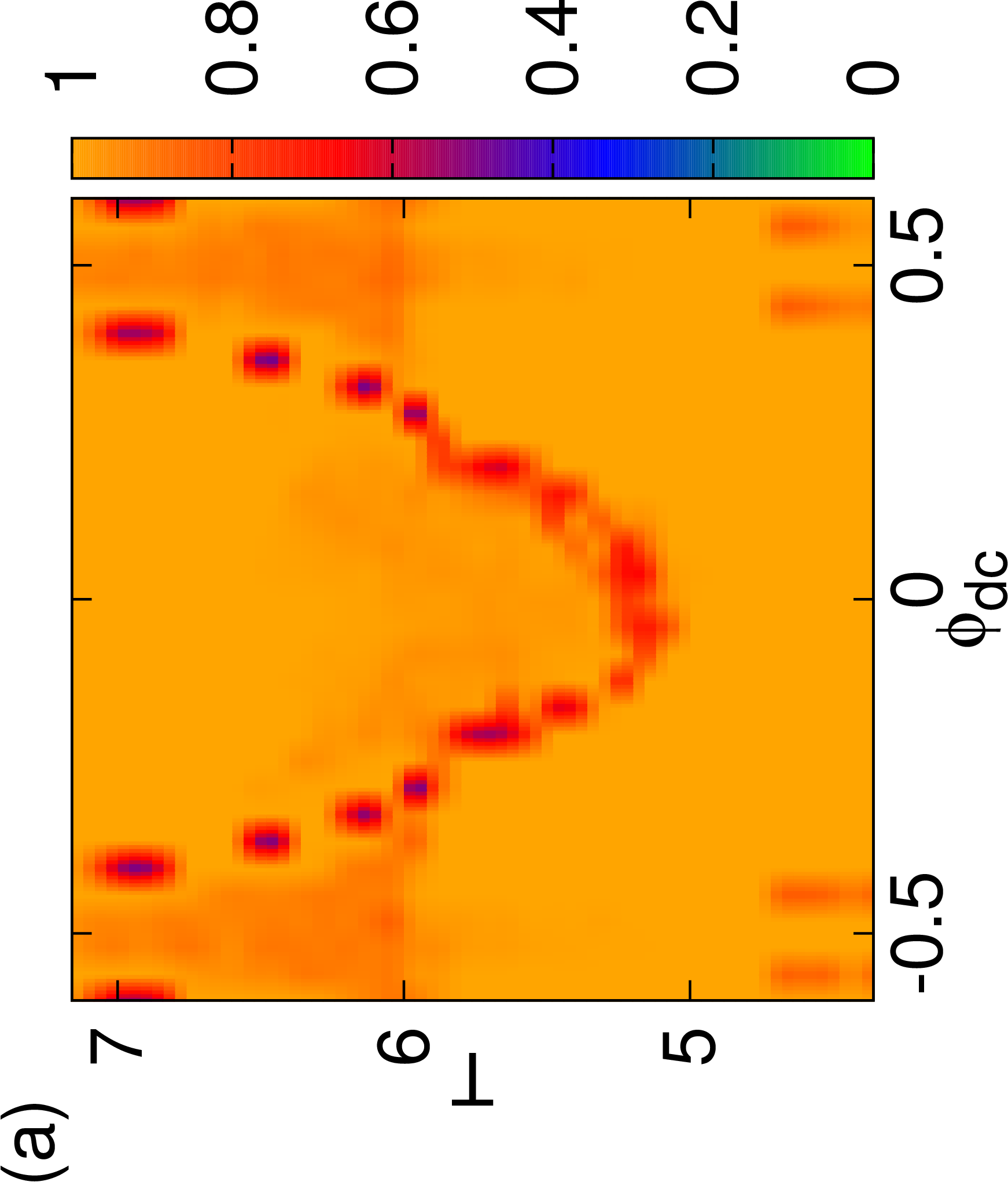}
\includegraphics[clip,width=0.28\textwidth,angle=-90]{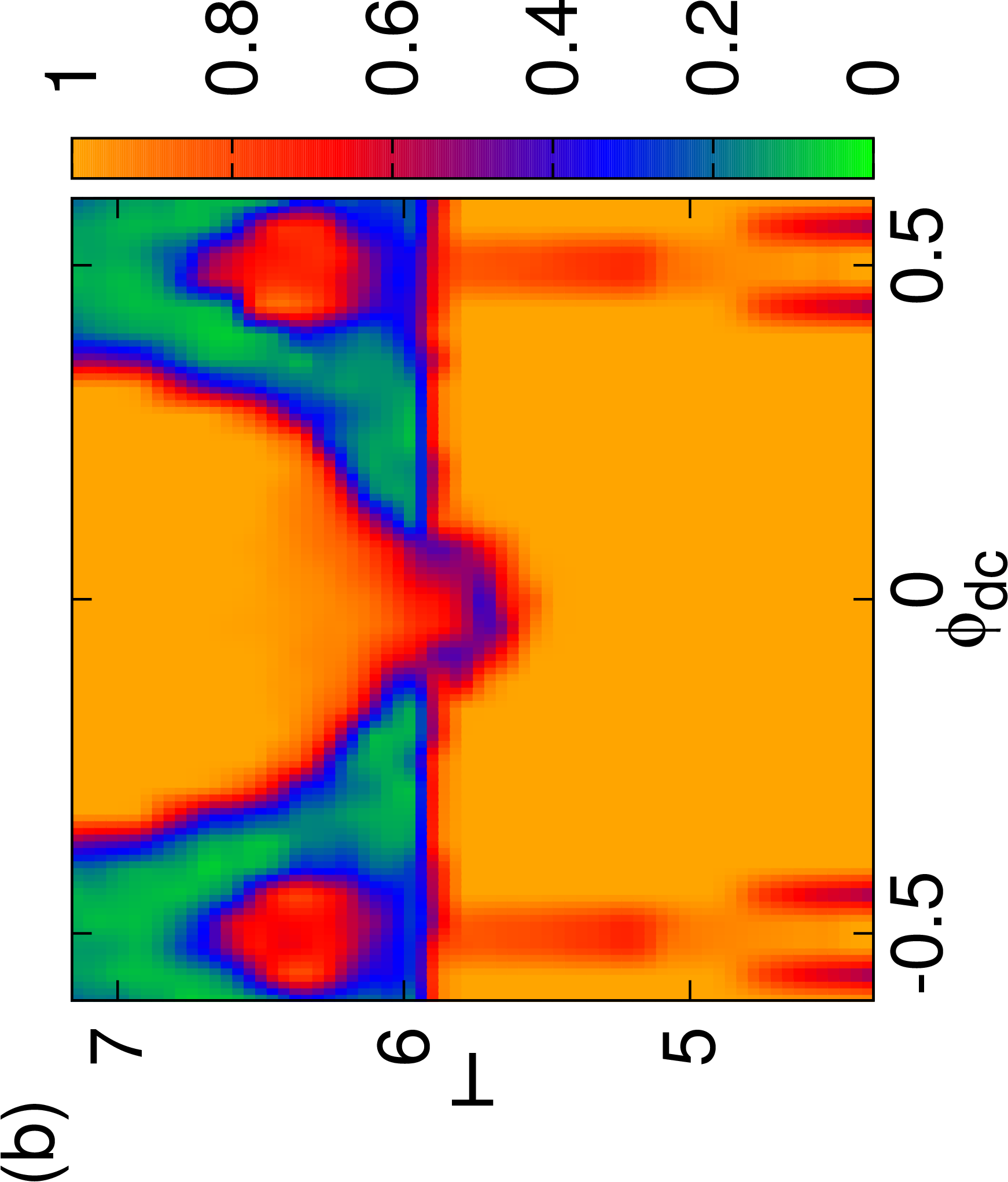}
\includegraphics[clip,width=0.28\textwidth,angle=-90]{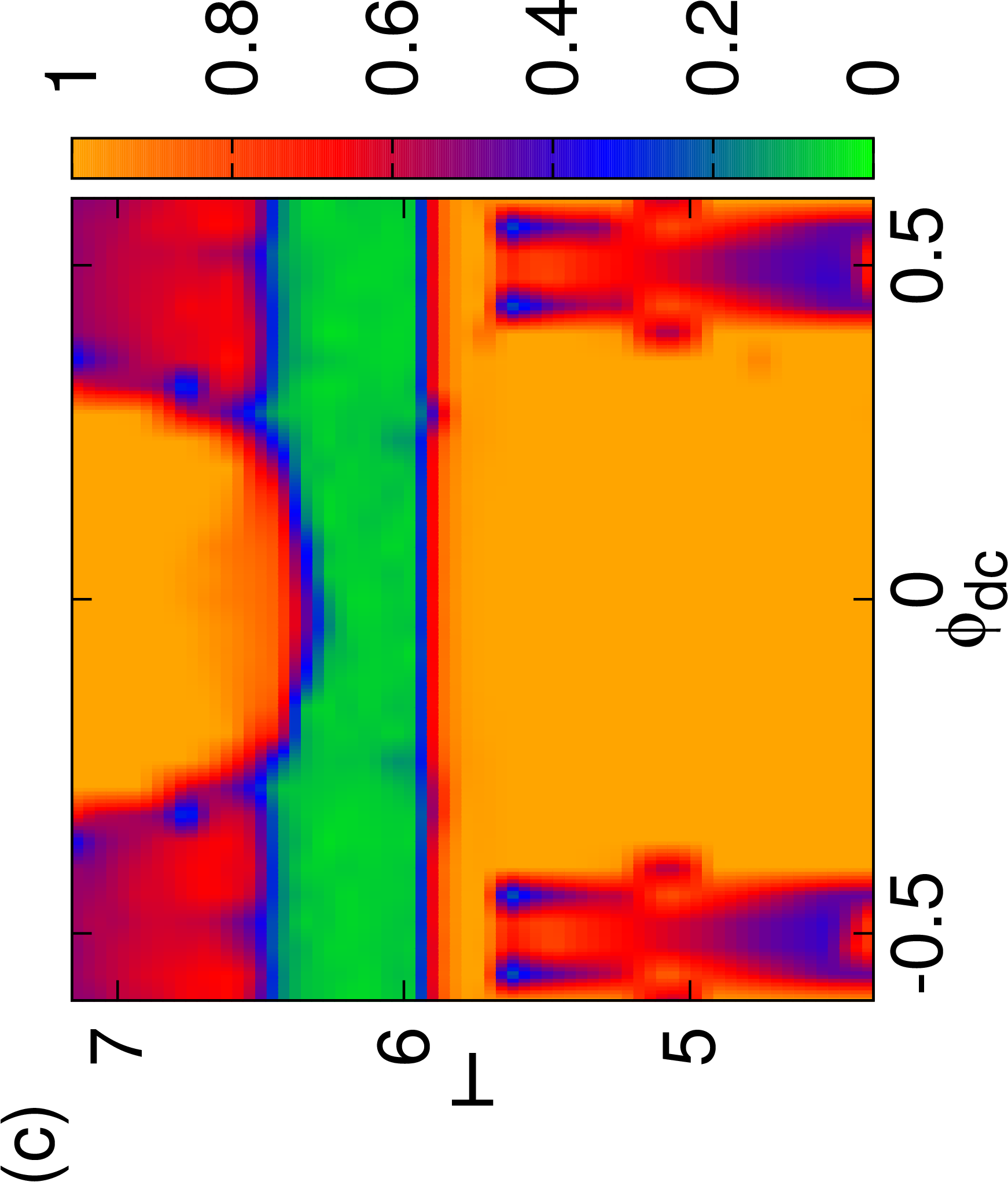}
\caption{(Color online) 
The magnitude of the synchronization parameter (Eq.~\ref{eq:globord}) in the parameter 
space $(T,\phi_{dc})$. where the brackets indicate averaging over the driving period 
$T=2\pi/\Omega$, with $\Omega$ being the normalized driving frequency. 
(a) $\phi_{ac}=0.003$, (b) $\phi_{ac}=0.015$, and (c) $\phi_ac=0.05$.
Other parameters as in Fig.~\ref{fig2}.
\label{fig6}}
\end{figure*}

In the previous sections we have seen how chimera states occur as a result of the high
multistability in the SQUID metamaterial. Moreover, it has been demonstrated that the external 
driving field dominates the frequency of the oscillations and therefore the level of 
synchronization in the entire device. The external driving field has three control parameters, 
which would be easily accessible in an experiment: the dc ($\phi_{dc}$) and ac ($\phi_{ac}$)
flux components and the driving period $T$ or equivalently the driving frequency $\Omega$. 
In Fig.~\ref{fig6} the magnitude of the 
global order parameter (Eq.~\ref{eq:globord}) averaged over the driving period 
$T=2\pi/\Omega$ is shown, in the parameter space of the $\phi_{dc}$ and $T$, for three
different values of $\phi_{ac}$. For very low values of ac component (Fig.~\ref{fig6}(a)) 
the level of synchronization is relatively high in the whole parameter space with distinct 
spots of lower coherence illustrated by their corresponding red color. Increasing $\phi_{ac}$ 
to the value used in the previous sections (Fig.~\ref{fig6}(b)), we observe that for 
$T<6$ the device is almost always synchronous. Above this critical value an area of 
desynchronization is formed, including within it a smaller ``island'' of high synchronization.
The blue colored thin areas mark the border between synchronization and desynchronization.
Note that the values for which chimera states were found in the previous sections ($T=5.9$ 
and $\phi_{dc}=0$) the system finds itself in an area very close to this transition, which 
intuitively is logical. For further increase of the ac component (Fig.~\ref{fig6}(c))
the synchronization areas shrink while the regions with less and lower coherence become 
larger. This is interesting because it shows that even though the driving field is 
stronger, it doesn't necessarily force all the SQUID oscillators to synchronize.
In any case, the order parameter cannot be zero, except in the thermodynamic limit;
for a finite system such as the SQUID metamaterial considered here, the order parameter
will reach a lower bound in the desynchronized state that depends on the size of the 
system $N$. Density plots like those shown in Fig.~\ref{fig6} may be useful for the 
construction of devices like parametric amplifiers that incorporate SQUID metamaterials
\cite{Castellanos2007,Castellanos2008} for which it is essential to avoid all types of 
noise, chaos, etc.
%%%%%%%%%%%%%%%%%%%%%%%%%%%%%%%%%%%%%%%%%%%%%%%%%%%%%%%%%%%%%%%%%%%%%%%%%%%%%%%%%%%%%%%%%%
%%%%%%%%%%%%%%%%%%%%%%%%%%%%%%%%%%%%%%%%%%%%%%%%%%%%%%%%%%%%%%%%%%%%%%%%%%%%%%%%%%%%%%%%%%

%%%%%%%%%%%%%%%%%%%%%%%%%%%%%%%%%%%%%%%%%%%%%%%%%%%%%%%%%%%%%%%%%%%%%%%%%%%%%%%%%%%%%%%%%%
\section{Conclusions}
\label{sec:sec5}
It has been demonstrated that single and double-headed chimera states coexist with 
solitary states and patterns of travelling incoherence in nonlocally coupled SQUID 
metamaterials. These coexisting patterns are a result of the high multistability of the 
individual SQUID oscillator which tenders the dynamics even more complex when we consider
an array of coupled elements. The spatial coherence was calculated through a local order 
parameter which reveals that the incoherent cluster periodically becomes coherent,
a sign of metastability previously studied.

Apart from the spatial complexity with coexisting high and low-flux amplitude oscillators,
the dynamics of the coupled system exhibit also temporal complexity. This is demonstrated 
through the phase portraits and time series of the individual SQUID oscillators, which
present a highly modulated quasiperiodic behavior. Evidence of chaos requires further 
studies which will be included in a future publication. Finally, by scanning the control 
parameter space of the external dc driving field and the driving period for various values of 
the ac field amplitude, we were able to locate the regions  where the transition from 
coherence to incoherence occurs. This is relevant for locating chimera states in the 
available parameter space which may also be useful in device applications.

Our numerical simulations rely on a realistic model, which is capable of reproducing 
several experimental findings such as the tunability patterns of two-dimensional SQUID 
metamaterials which are obtained by varying an applied dc flux and the driving frequency
\cite{Tsironis2014b}. Since several experiments in one- and two-dimensional SQUID 
metamaterials have been already carried out, revealing very interesting properties peculiar
to them, we believe that chimera states could in principle be detected with the presently
available set-ups. Thus, SQUID metamaterials may serve as a prototype, test-bed system
to check and/or confirm theories and concepts of nonlinear dynamics of coupled oscillators.
\newline
%%%%%%%%%%%%%%%%%%%%%%%%%%%%%%%%%%%%%%%%%%%%%%%%%%%%%%%%%%%%%%%%%%%%%%%%%%%%%%%%%%%%%%%%%%

%%%%%%%%%%%%%%%%%%%%%%%%%%%%%%%%%%%%%%%%%%%%%%%%%%%%%%%%%%%%%%%%%%%%%%%%%%%%%%%%%%%%%%%%%%
{\bf Acknowledgements}

This work was partially supported by
the European Union Seventh Framework Programme (FP7-REGPOT-2012-2013-1)
under grant agreement no 316165,
the Ministry of Education and Science of the Russian Federation in the
framework of the Increase Competitiveness Program of NUST ``MISiS'' (No. K2-2015-007),
and the Thales Project MACOMSYS, cofinanced by the European Union (European
Social Fund ESF) and Greek national funds through the
Operational Program ``Education and Lifelong Learning'' of
the National Strategic Reference Framework (NSRF) Research
Funding Program: THALES Investing in Knowledge Society
via the European Social Fund.

%%%%%%%%%%%%%%%%%%%%%%%%%%%%%%%%%%%%%%%%%%%%%%%%%%%%%%%%%%%%%%%%%%%%%%%%%%%%%%%%%%%%%%%%%%

%%%%%%%%%%%%%%%%%%%%%%%%%%%%%%%%%%%%%%%%%%%%%%%%%%%%%%%%%%%%%%%%%%%%%%%%%%%%%%%%%%%%%%%%%%

\end{document}